\documentclass[aps, prd, amsmath, floats, floatfix, twocolumn, nofootinbib, nosuperscriptaddress]{revtex4-1}
\usepackage[pdftex]{graphicx}
\usepackage{color}
\usepackage{latexsym}
\usepackage[colorlinks]{hyperref}

\newcommand{\be}{\begin{eqnarray}}
\newcommand{\ee}{\end{eqnarray}}

\begin{document}

\title{Search for traversable wormholes in active galactic nuclei using X-ray data}

\author{Ashutosh~Tripathi}
\affiliation{Center for Field Theory and Particle Physics and Department of Physics, Fudan University, 200438 Shanghai, China}

\author{Biao~Zhou}
\affiliation{Center for Field Theory and Particle Physics and Department of Physics, Fudan University, 200438 Shanghai, China}

\author{Askar~B.~Abdikamalov}
\affiliation{Center for Field Theory and Particle Physics and Department of Physics, Fudan University, 200438 Shanghai, China}

\author{Dimitry~Ayzenberg}
\affiliation{Center for Field Theory and Particle Physics and Department of Physics, Fudan University, 200438 Shanghai, China}

\author{Cosimo~Bambi}
\email[Corresponding author: ]{bambi@fudan.edu.cn}
\affiliation{Center for Field Theory and Particle Physics and Department of Physics, Fudan University, 200438 Shanghai, China}

\begin{abstract}
In a previous paper, one of us calculated iron K$\alpha$ line profiles emitted from possible accretion disks around traversable wormholes as a first step to use X-ray reflection spectroscopy to search for astrophysical wormholes in active galactic nuclei. In the present paper, we extend that work and construct an XSPEC model for the whole relativistic reflection spectrum. We apply our model to \textsl{XMM-Newton} and \textsl{NuSTAR} observations of the supermassive object in MCG--6--30--15 and we check whether these observations prefer the hypothesis that the central body is a Kerr black hole or a traversable wormhole. We find that the two models provide equally good fits, so with the available data we cannot distinguish the black hole and wormhole scenarios. 
\end{abstract}

\maketitle


\section{Introduction}

Wormholes are hypothetical topological structures connecting either two faraway regions in the same universe with a shortcut or two different universes that are otherwise disconnected~\cite{wh1,wh2,wh3}. They are not really a prediction of the theory of general relativity, but rather a generic possibility in curved spacetimes. Traversable wormholes are particularly interesting because they would be ``traversable'': one can go through the wormhole along one direction and then go back along the opposite direction, so the wormhole could be effectively used as a shortcut to reach regions otherwise too far away to visit. Wormholes may have been created in the early Universe, when topological transitions were possible due to quantum gravity effects, and have somehow survived until today, or they may form in the contemporary Universe as a consequence of the formation of a baby-universe inside our Universe.

In the past years, a number of authors have studied possible observational phenomena to identify astrophysical wormholes in our Universe~\cite{p1,p2,p3,p4,p5,p6,p7,p8,p9,p10,p11,p12,p13}. At large distances, most wormholes look like compact objects and thus may mimic black holes. For example, the supermassive objects with masses of millions or billions of Solar masses at the center of every normal galaxy are normally interpreted as supermassive black holes. However, the origin of these objects is still puzzling: we know supermassive objects with masses of billions of Solar masses at redshift $z \gtrsim 6$~\cite{z1,z2,z3}, less than 1~Gyr after the big bang, and we do not understand how black holes created from gravitational collapse of primordial stars could have become so heavy in such a short time. So there is the possibility that they are actually primordial relics of the very early Universe~\cite{dolgov}, and they may be wormholes rather than black holes. The black hole and wormhole scenarios can be tested by probing the spacetime metric around these objects.

There are a number of available and proposed methods for testing the nature of the supermassive objects in galactic nuclei with electromagnetic techniques~\cite{review}, while gravitational wave tests will only be possible with future gravitational wave space antennas. Among the electromagnetic techniques, X-ray reflection spectroscopy is surely the most mature one and the only method that has been already successfully applied to a number of sources, showing that it is able to provide interesting constraints~\cite{n1,n1b,n2,n3,n4,n5,n6}.

In Ref.~\cite{p6}, one of us studied the shape of the iron line expected from a thin accretion disk around a particular class of traversable wormholes. A broad iron line is usually the most prominent feature of the reflection spectrum of an accretion disk around a compact object, but any attempt to fit observational data and measure the properties of the system should involve the analysis of the whole reflection spectrum, not only of the iron line. Within such a simplified framework, the conclusion was that the iron line from accretion disks around non-rotating or very slow-rotating ($a_* \lesssim 0.02$, where $a_*$ is the dimensionless spin parameter) wormholes could mimic that expected from fast-rotating Kerr black holes, while even for moderate values of the spin parameter of the wormhole the iron line shape was quite different from that of black holes and thus inconsistent with the iron lines observed from the supermassive objects in galactic nuclei.

In the present paper, we extend the study in Ref.~\cite{p6} and construct a model for the whole reflection spectrum produced by an accretion disk around a wormhole. Then we consider the supermassive object in the nucleus of the galaxy MCG--6--30--15. This is a bright source, characterized by a prominent and relativistically broadened iron line, and was simultaneously observed by \textsl{XMM-Newton} and \textsl{NuSTAR} in 2013~\cite{andrea}. The reflection spectrum of the source is fitted with a Kerr model and our wormhole model to check which one can fit the data better and if observations can select one of the two models and rule out the other one. Our results nicely confirm the preliminary conclusions of Ref.~\cite{p6}. With the available data, we cannot distinguish the hypothesis that the supermassive object in MCG--6--30--15 is a black hole from that in which it would be a non-rotating or very slow-rotating traversable wormhole. The quality of the best fit of the two models is quite similar.

The content of the paper is as follows. In the next section, we present our reflection model for traversable wormholes. In Section~\ref{s-3}, we apply the new model to the \textsl{XMM-Newton} and \textsl{NuSTAR} data of MCG--6--30--15 of 2013. In Section~\ref{s-4}, we discuss our results and present our conclusions. The details on the \textsl{XMM-Newton} and \textsl{NuSTAR} observations, data reduction, and data analysis are reported in Appendix~\ref{s-aaa}. Throughout the paper, we use natural units in which $c = G_{\rm N} = 1$ and a metric with signature $(-+++)$.

\section{Wormhole reflection model \label{s-2}}

X-ray reflection spectroscopy refers to the analysis of the reflection spectrum of an accretion disk of a compact object~\cite{ref1,ref2,ref3,ref4}, which is normally a neutron star or a black hole, but could potentially be an even more exotic object like a wormhole. For a mass accretion rate ranging between a few percent and up to about 30\% of the Eddington limit of the source, the accretion disk is geometrically thin and optically thick. Every point on the surface of the accretion disk has a blackbody-like spectrum, so the whole disk turns out to have a multi-temperature blackbody spectrum. Thermal photons from the disk can inverse Compton scatter off free electrons in the ``corona'', which is a generic name to indicate some hot, usually compact and optically thin, material near the compact object. The corona may be the base of the jet, the atmosphere above the accretion disk, the accretion flow in the plunging region between the inner edge of the disk and the compact object, etc. The Comptonized photons have a power-law spectrum with a high energy cutoff. These photons can illuminate the accretion disk, producing a reflection component~\cite{peda}.

In the rest-frame of the gas, the reflection spectrum is characterized by some fluorescent emission lines in the soft X-ray band ($< 7$~keV) and a Compton hump with a peak at 20-30~keV. The most prominent feature of the reflection spectrum is often the iron K$\alpha$ complex, which is around 6.4~keV in the case of neutral or weakly ionized iron and shifts up to 6.97~keV in the case of H-like ions. Due to Doppler boosting and gravitational redshift, which are different at different points on the accretion disk, the fluorescent emission lines of the reflection spectrum become very broad for an observer far from the source. In the presence of the correct astrophysical model and of high quality data, X-ray reflection spectroscopy can be a powerful tool to probe the spacetime metric around the compact object~\cite{i1,i2,i3,i4}.

We construct a relativistic reflection model for traversable wormholes implementing the wormhole metric in the reflection model {\sc relxill\_nk}~\cite{nk1,nk2}. As in Ref.~\cite{p6}, we consider the class of traversable wormholes described by the following line element~\cite{teo,p4}
\be\label{eq-metric}
ds^2 &=& - e^{2\Phi} dt^2 + \frac{dr^2}{1 - \left(\frac{r_0}{r}\right)^\gamma} \nonumber\\
&& + r^2 \left[ d\theta^2 + \sin^2\theta \left( d\phi - \omega dt \right) \right] \, ,
\ee
where $\Phi = - r_0/r$ is the redshift function, $\omega = 2J/r^3$, $r_0$ is the wormhole throat setting the scale of the system, $J$ is the wormhole spin angular momentum, and $\gamma$ is a constant and corresponds to the parameter describing the wormhole family.

The reflection spectrum of the accretion disk as seen by a distant observer can be written as
\be
F_{\rm o} \left( \nu_{\rm o} \right) &=& \int I_{\rm o} \left( \nu_{\rm o} , X , Y \right) \, d\Omega 
\nonumber\\
&=& \int g^3 I_{\rm e} \left( \nu_{\rm e} , r_{\rm e} , \vartheta_{\rm e} \right) \, d\Omega \, ,
\ee 
where $I_{\rm o}$ and $I_{\rm e}$ are the specific intensity of the radiation as measured, respectively, by the distant observer and in the rest-frame of the gas, $X$ and $Y$ are the Cartesian coordinates of the disk image in the plane of the distant observer, $d\Omega = dXdY/D^2$ is the element of the solid angle subtended by the disk image in the observer's sky, $D$ is the distance of the observer from the source, and $r_{\rm e}$ and $\vartheta_{\rm e}$ are, respectively, the emission radius in the disk and the emission angle. $I_{\rm o} = g^3 I_{\rm e}$ follows from Liouville's theorem, where $g = \nu_{\rm o}/\nu_{\rm e}$ is the redshift factor, $\nu_{\rm o}$ is the photon frequency measured by the distant observer, and $\nu_{\rm e}$ is the photon frequency measured in the rest-frame of the gas in the disk.

Employing the formalism of the transfer function proposed by Cunningham~\cite{cun}, the observed flux can be written as
\be
F_{\rm o} \left( \nu_{\rm o} \right) &=& \frac{1}{D^2} \int_{r_{\rm in}}^{r_{\rm out}} \int_0^1
\pi r_{\rm e} \frac{g^2}{\sqrt{g^* \left(1 - g^*\right)}} \nonumber\\
&& \quad \times f \left( \nu_{\rm e} , r_{\rm e} , \vartheta_{\rm e} \right) \, 
I_{\rm e} \left( \nu_{\rm e} , r_{\rm e} , \vartheta_{\rm e} \right) \, dg^* \, dr_{\rm e} \, ,
\ee
where $r_{\rm in}$ and $r_{\rm out}$ are, respectively, the inner and the outer edges of the accretion disk, $f$ is the ``transfer function'' defined as
\be
f \left( \nu_{\rm e} , r_{\rm e} , \vartheta_{\rm e} \right) = 
\frac{g \sqrt{g^* \left(1 - g^*\right)}}{\pi r_{\rm e}} 
\left| \frac{\partial \left( X , Y \right)}{\partial \left( g^* , r_{\rm e} \right)} \right| \, ,
\ee
where $| \partial \left( X , Y \right) / \partial \left( g^* , r_{\rm e} \right) |$ is the Jacobian between the Cartesian coordinates of the distant observer and the disk coordinates $g^*$ and $r_{\rm e}$, and $g^* = g^* \left( r_{\rm e} , \vartheta_{\rm e} \right)$ is the relative redshift factor defined as
\be
g^* = \frac{g - g_{\rm min}}{g_{\rm max} - g_{\rm min}} \, 
\ee
and ranges from 0 to 1. $g^*_{\rm max} = g^*_{\rm max} \left( r_{\rm e} , \vartheta_{\rm e} \right)$ and $g^*_{\rm min} = g^*_{\rm min} \left( r_{\rm e} , \vartheta_{\rm e} \right)$ are, respectively, the maximum and the minimum values of the redshift factor $g$ for the photons emitted from the radial coordinate $r_{\rm e}$ and detected by a distant observer at the viewing angle $i$.

\begin{figure}[t]
\begin{center}
\includegraphics[type=pdf,ext=.pdf,read=.pdf,width=8.5cm]{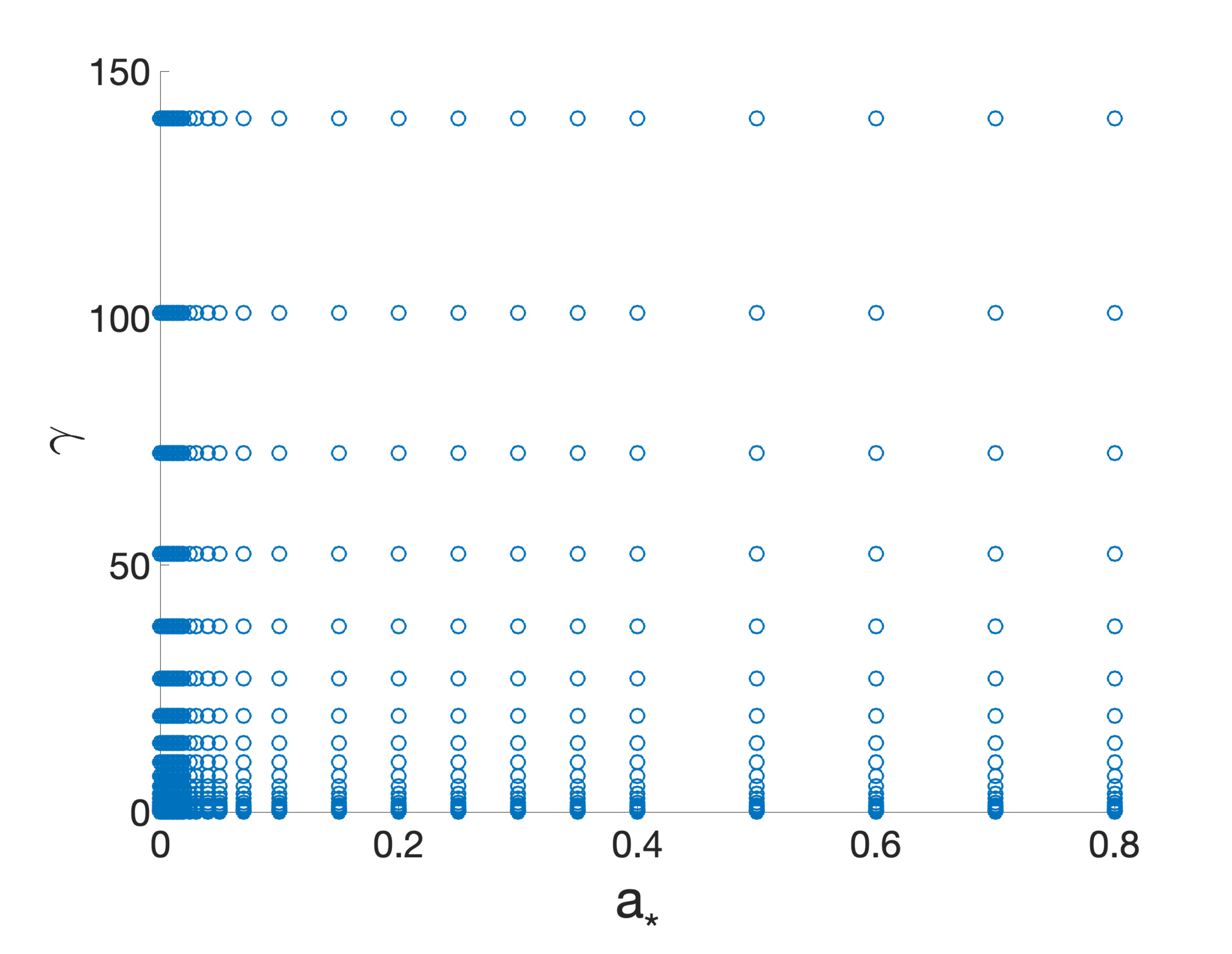}
\end{center}
\vspace{-0.5cm}
\caption{Grid points in the FITS file of the wormhole model for the spin parameter $a_*$ and the parameter $\gamma$. See the text for more details. \label{f-grid}}
\end{figure}

\begin{figure*}[t]
\begin{center}
\vspace{-0.6cm}
\includegraphics[type=pdf,ext=.pdf,read=.pdf,width=8.0cm,trim={0.5cm 0 3cm 11cm},clip]{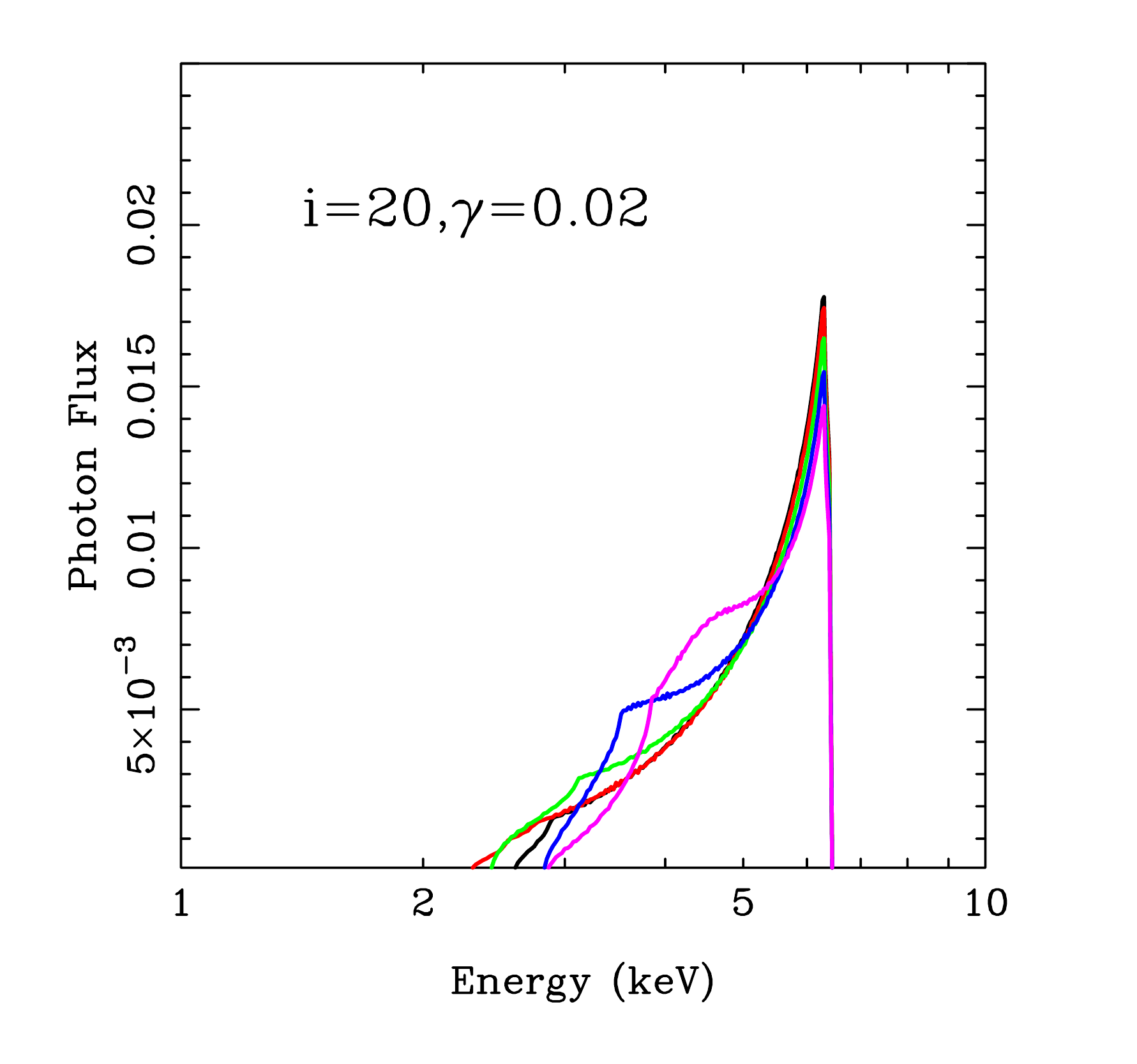}
\includegraphics[type=pdf,ext=.pdf,read=.pdf,width=8.0cm,trim={0.5cm 0 3cm 11cm},clip]{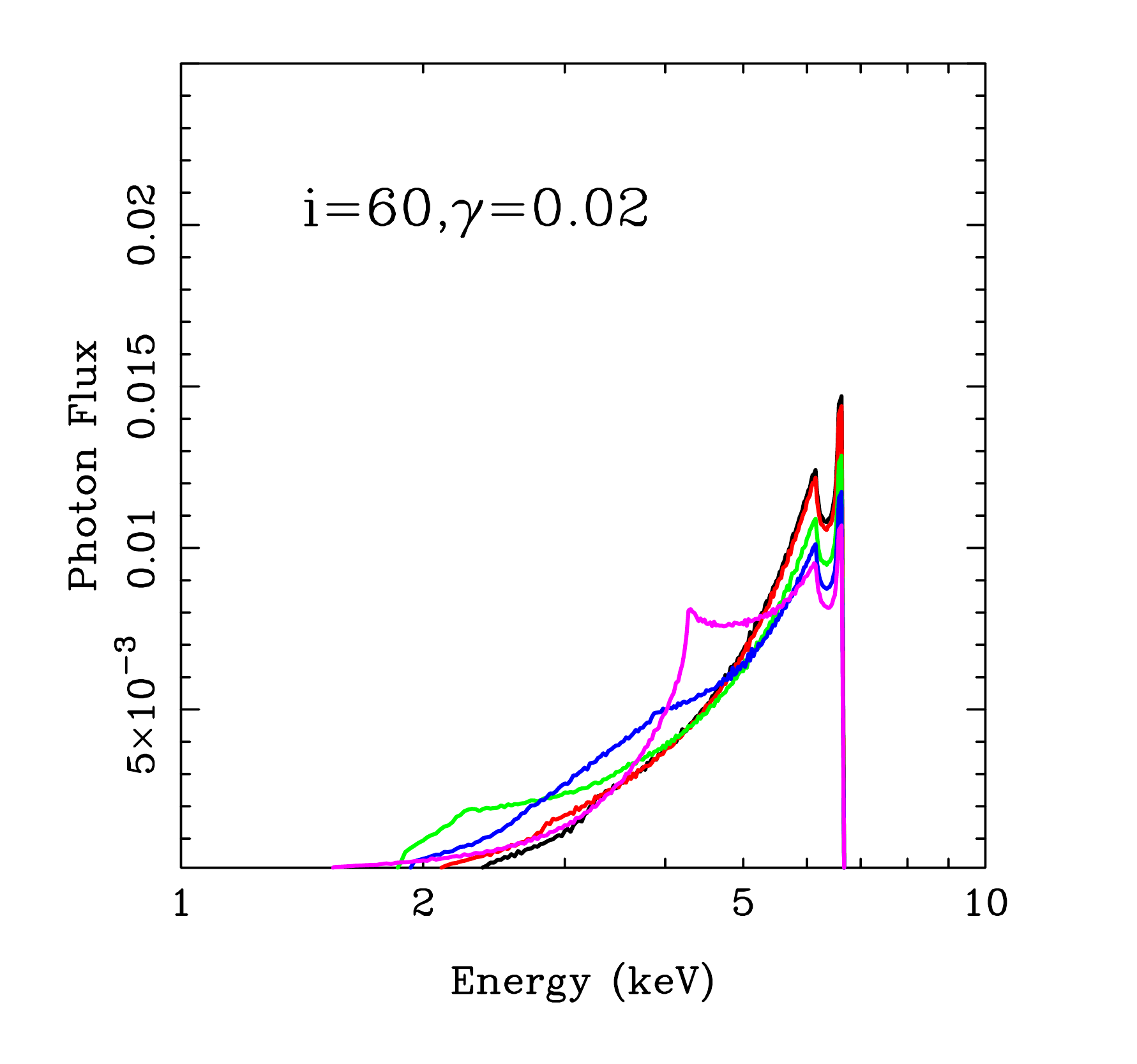}\\
\vspace{-0.2cm}
\includegraphics[type=pdf,ext=.pdf,read=.pdf,width=8.0cm,trim={0.5cm 0 3cm 11cm},clip]{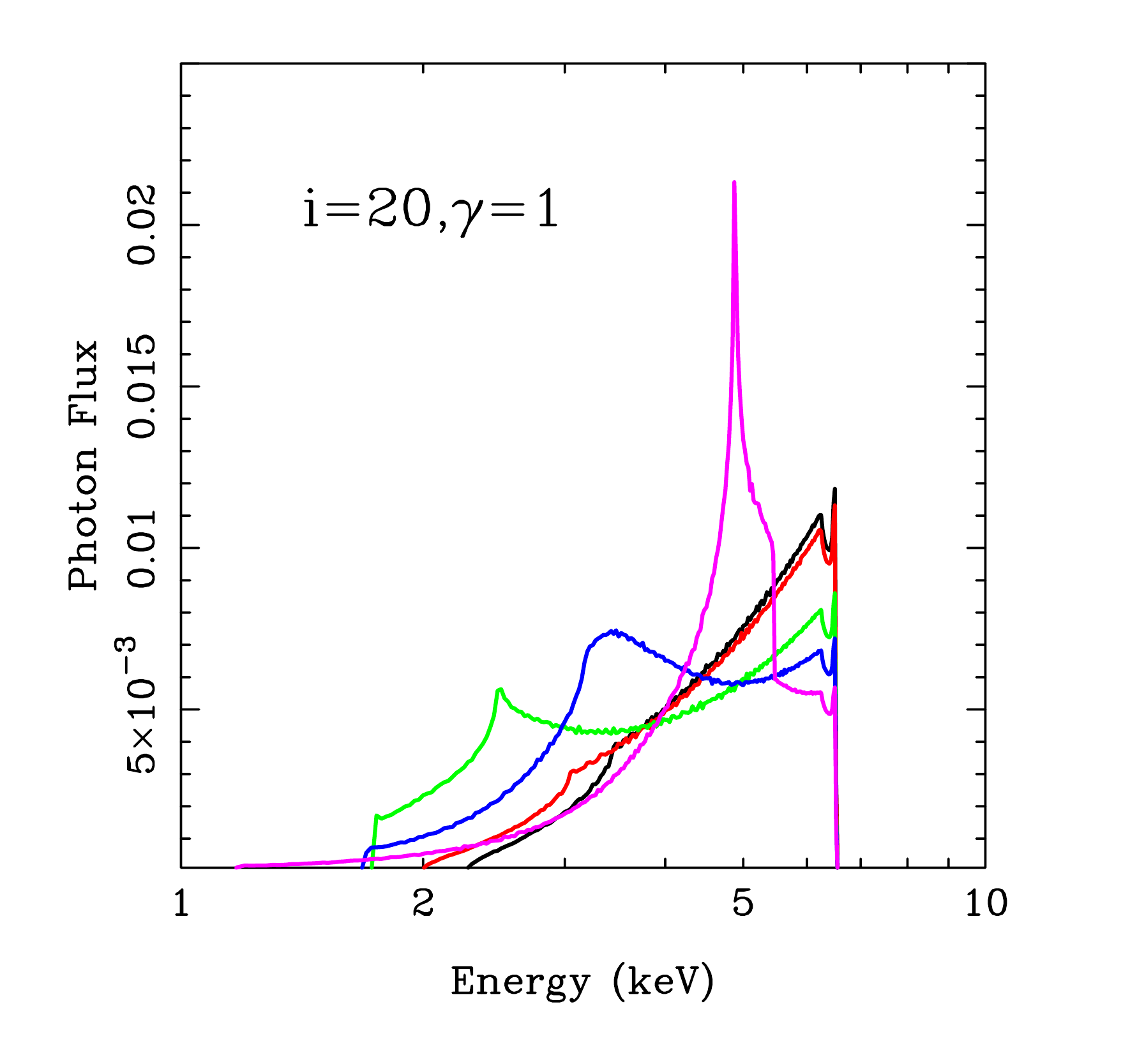}
\includegraphics[type=pdf,ext=.pdf,read=.pdf,width=8.0cm,trim={0.5cm 0 3cm 11cm},clip]{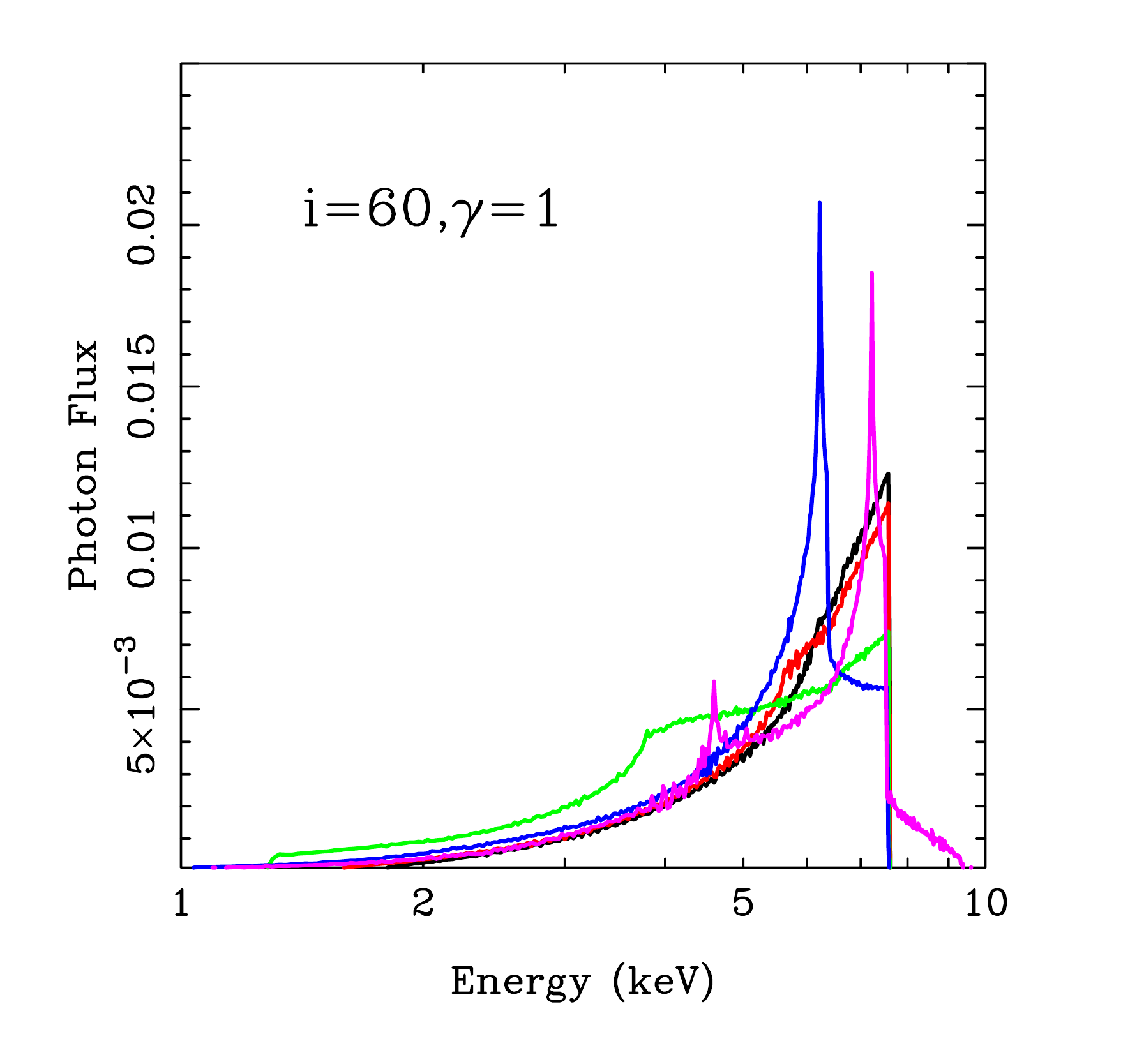}\\
\vspace{-0.2cm}
\includegraphics[type=pdf,ext=.pdf,read=.pdf,width=8.0cm,trim={0.5cm 0 3cm 11cm},clip]{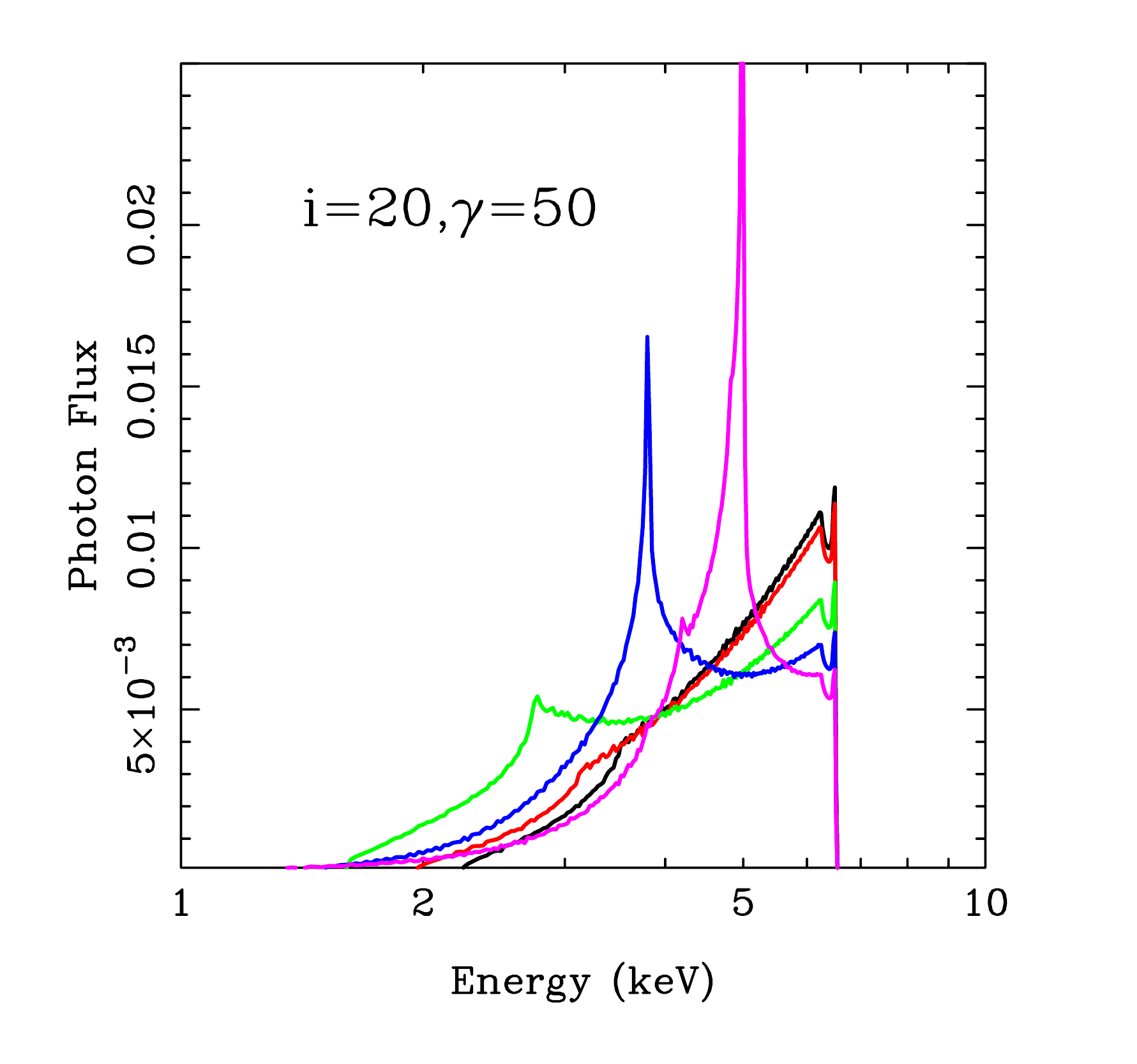}
\includegraphics[type=pdf,ext=.pdf,read=.pdf,width=8.0cm,trim={0.5cm 0 3cm 11cm},clip]{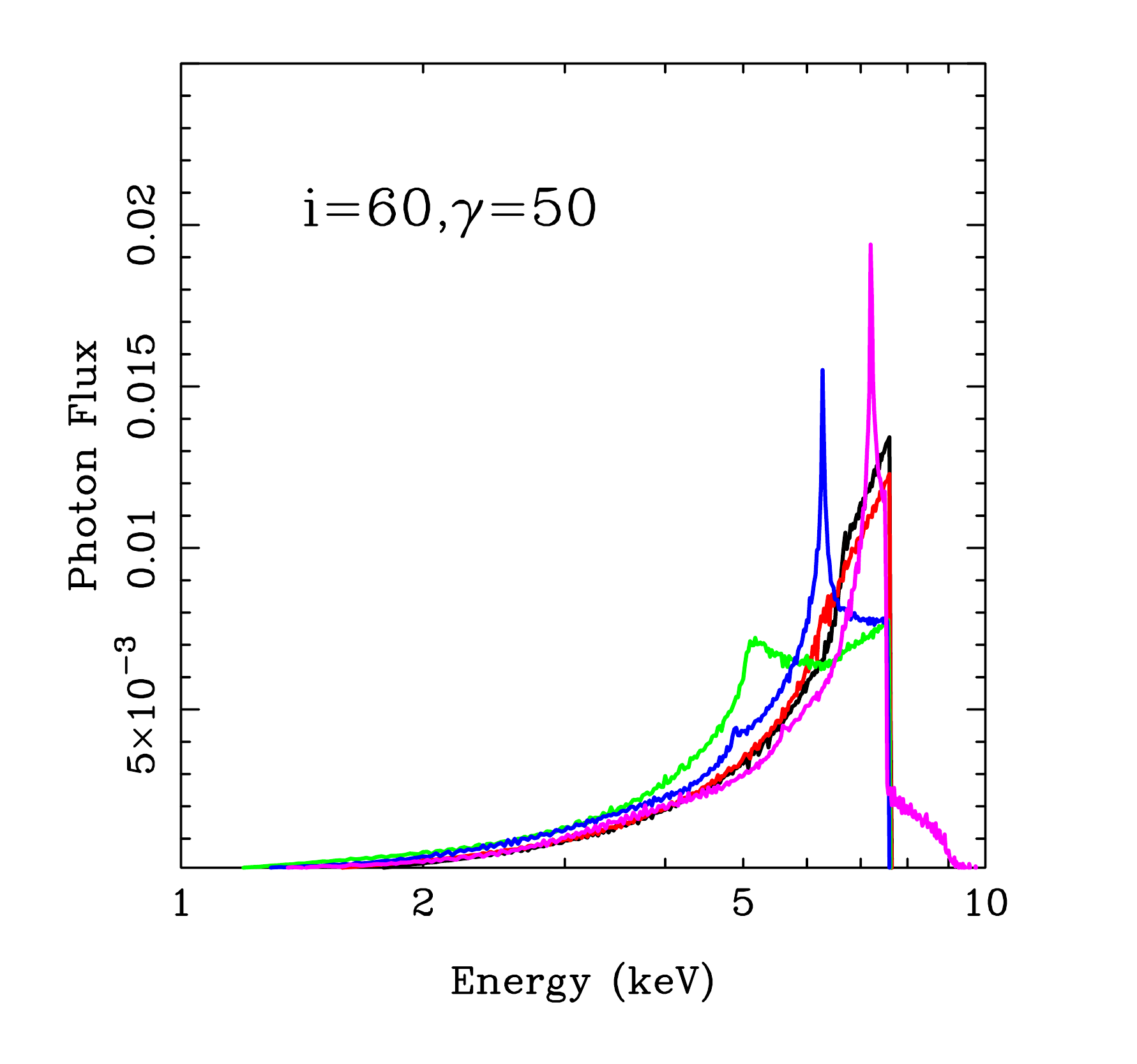}
\end{center}
\vspace{-0.7cm}
\caption{Iron line profiles in wormhole spacetime. The viewing angle of the observer is $i = 20^\circ$ (left panel) and $i = 60^\circ$ (right panel). The value of the spin parameter is $a_* = 0$ (black line), 0.01 (red line), 0.1 (green line), 0.3 (blue line), and 0.8 (magenta line). The emissivity index of the intensity profile is always $q = 3$. \label{f-w}}
\end{figure*}

\begin{figure*}[t]
\begin{center}
\vspace{-0.6cm}
\includegraphics[type=pdf,ext=.pdf,read=.pdf,width=8.0cm,trim={0.5cm 0 3cm 11cm},clip]{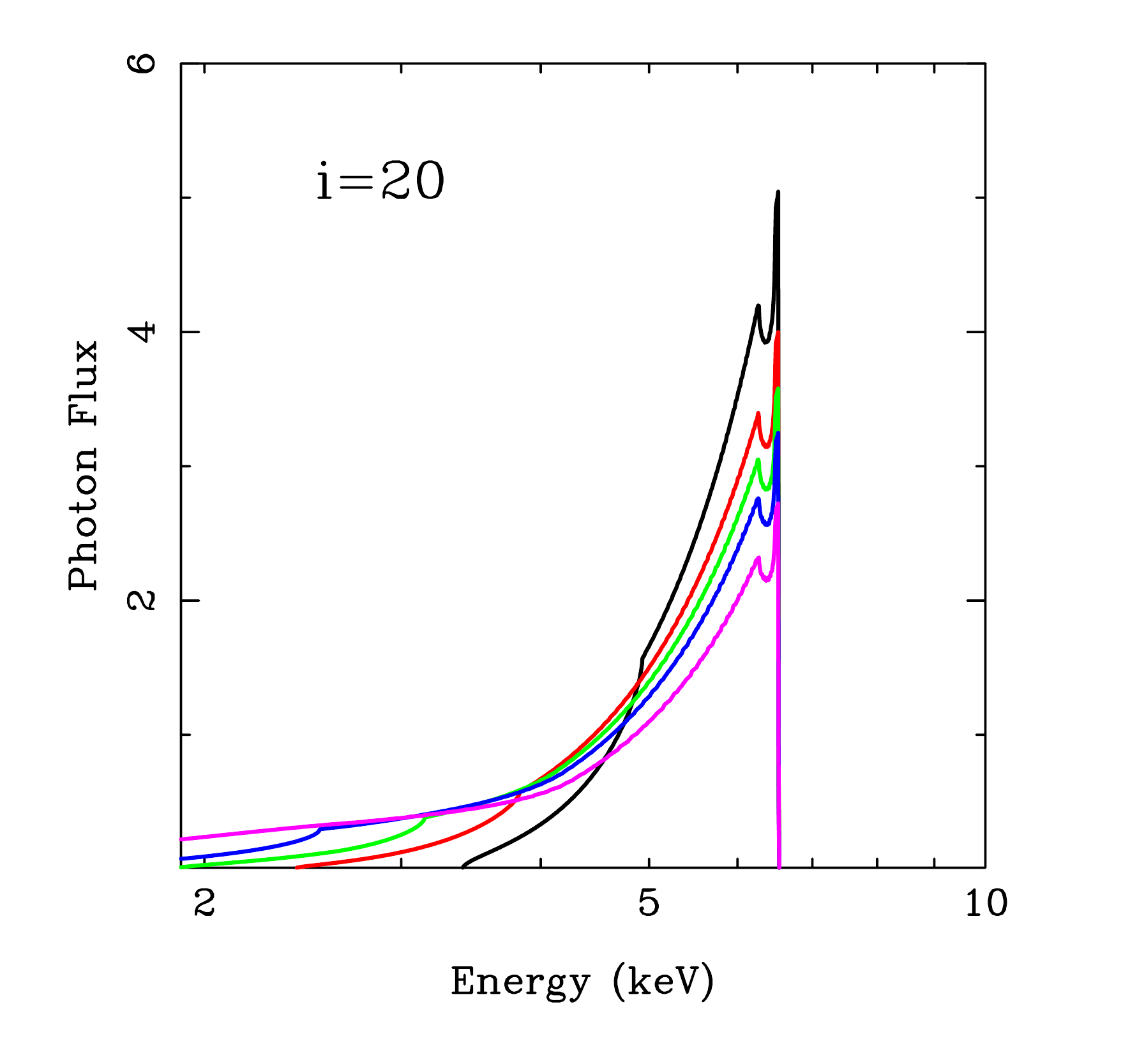}
\includegraphics[type=pdf,ext=.pdf,read=.pdf,width=8.0cm,trim={0.5cm 0 3cm 11cm},clip]{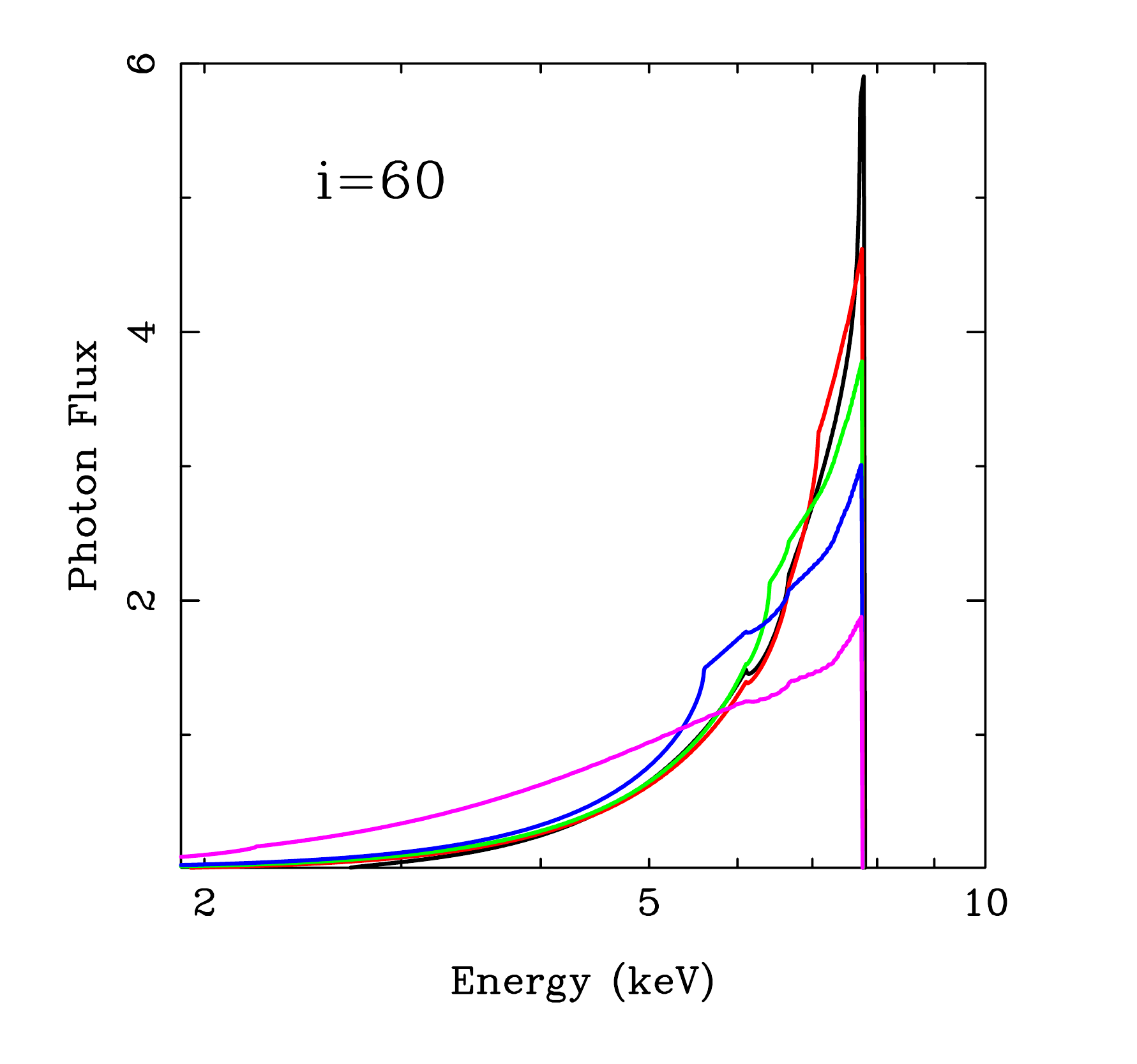}
\end{center}
\vspace{-0.7cm}
\caption{Iron line profiles in Kerr spacetime. The viewing angle of the observer is $i = 20^\circ$ (left panel) and $i = 60^\circ$ (right panel). The value of the spin parameter is $a_* = 0.4$ (black line), 0.8 (red line), 0.9 (green line), 0.95 (blue line), and 0.998 (magenta line). The emissivity index of the intensity profile is always $q = 3$. \label{f-k}}
\end{figure*}

The transfer function only depends on the spacetime metric and the viewing angle of the distant observer, and takes all relativistic effects into account (Doppler boosting, gravitational redshift, light bending). More details can be found in Ref.~\cite{nk1}. Here we use a ray-tracing code to calculate the Jacobian $| \partial \left( X , Y \right) / \partial \left( g^* , r_{\rm e} \right) |$ and then the transfer function in the wormhole spacetime~(\ref{eq-metric}). We calculate the transfer function for a grid of points in the 3D space $( a_* , \gamma , i)$, where $a_* = J/r_0^2$ is the dimensionless spin parameter of the wormhole. We use the same choice as in Ref.~\cite{nk1} for the points of the viewing angles in the grid, while the points on the plane $( a_* , \gamma )$ are shown in Fig.~\ref{f-grid}. We tabulate the transfer function in a FITS file and we can then use the {\sc relxill\_nk} package (with the exception of the flavor for lamppost coronal geometry, which requires additional metric-dependent calculations) for the calculations of the single line profile and full reflection spectra~\cite{nk1,nk2}. Examples of iron line profiles in the traversable wormhole spacetime calculated using our model are shown in Fig.~\ref{f-w}, and they can be compared with the iron line profiles in Kerr spacetime reported in Fig.~\ref{f-k}. In all plots, the emissivity index of the intensity profile is $q = 3$; that is, $I_{\rm e} \propto r_{\rm e}^{-3}$.

\begin{table*}
\centering
\scalebox{1.0}{
\begin{tabular}{l|cccc|cccc}
\hline\hline
Model & \multicolumn{4}{c}{Kerr metric} & \multicolumn{4}{c}{Wormhole metric} \\
\hline
Group & 1 & 2 & 3 & 4 & 1 & 2 & 3 & 4 \\
\hline
{\sc tbabs} &&&& &&&& \\
$N_{\rm H} / 10^{22}$ cm$^{-2}$ & \multicolumn{4}{c}{$0.039^\star$} & \multicolumn{4}{c}{$0.039^\star$} \\
\hline
{\sc warmabs$_1$} &&&& \\
$N_{\rm H \, 1} / 10^{22}$ cm$^{-2}$ & $0.49^{+0.15}_{-0.06}$ & $1.181^{+0.013}_{-0.066}$ & $1.007^{+0.013}_{-0.027}$ & $0.25^{+0.04}_{-0.11}$ 
& $0.640_{-0.039}^{+0.014}$ & $1.187_{-0.053}^{+0.018}$ & $1.022_{-0.029}^{+0.020}$ & $0.24_{-0.05}^{+0.06}$ \\
$\log\xi_1$ & $1.86^{+0.98}_{-0.04}$ & $1.954^{+0.009}_{-0.025}$ & $1.921^{+0.011}_{-0.031}$ & $2.48^{+0.31}_{-0.09}$ 
& $1.91_{-0.08}^{+0.03}$ & $1.955_{-0.020}^{+0.014}$ & $1.919_{-0.012}^{+0.016}$ & $2.48_{-0.16}^{+0.14}$ \\
\hline
{\sc warmabs$_2$} &&&& \\
$N_{\rm H \, 2} / 10^{22}$ cm$^{-2}$ & $0.62^{+0.04}_{-0.07}$ & -- & $0.55^{+0.18}_{-0.10}$ & $0.74^{+0.12}_{-0.03}$ 
& $0.48_{-0.17}^{+0.17}$ & -- & $0.61_{-0.13}^{+0.21}$ & $0.75_{-0.04}^{+0.07}$ \\
$\log\xi_2$ & $1.909^{+0.020}_{-0.114}$ & -- & $3.24^{+0.05}_{-0.05}$ & $1.829^{+0.010}_{-0.020}$ 
& $1.86_{-0.04}^{+0.04}$ & -- & $3.26_{-0.07}^{+0.05}$ & $1.827_{-0.023}^{+0.013}$ \\
\hline
{\sc dustyabs} &&&& \\
$\log \big( N_{\rm Fe} / 10^{21}$ cm$^{-2} \big)$ & \multicolumn{4}{c}{$17.409^{+0.021}_{-0.034}$} & \multicolumn{4}{c}{$17.408^{+0.020}_{-0.038}$} \\
\hline
{\sc cutoffpl} &&&& \\
$\Gamma$ & $1.951^{+0.010}_{-0.003}$ & $1.973^{+0.004}_{-0.010}$ & $2.013^{+0.005}_{-0.011}$ & $2.027^{+0.004}_{-0.011}$ 
& $1.949_{-0.005}^{+0.012}$ & $1.971_{-0.006}^{+0.013}$ & $2.020_{-0.009}^{+0.012}$ & $2.032_{-0.015}^{+0.007}$ \\
$E_{\rm cut}$ [keV] & $199^{+15}_{-26}$ & $156^{+15}_{-18}$ & $169^{+21}_{-20}$ & $291^{+141}_{-60}$
& $200_{-35}^{+35}$ & $155_{-21}^{+35}$ & $174_{-27}^{+39}$ & $309_{-75}^{+189}$ \\
$N_\text{\sc cutoffpl}$~$(10^{-3})$ & $8.39^{+0.08}_{-0.34}$ & $12.44^{+0.12}_{-0.35}$  & $15.25^{+0.17}_{-1.28}$ & $21.14^{+0.22}_{-1.78}$
& $8.4_{-0.3}^{+0.3}$ & $12.52_{-0.73}^{+0.23}$ & $16.5_{-1.2}^{+0.4}$ & $22.2_{-0.7}^{+0.5}$ \\ 
\hline
{\sc relxill\_nk} &&&& \\
$q_{\rm in}$ & $6.4^{+0.4}_{-1.3}$ & $6.7^{+0.4}_{-0.9}$ & $7.6^{+0.6}_{-0.5}$ & $8.4^{+0.6}_{-0.7}$
& $4.8_{-1.2}^{+1.2}$ & $3.8_{-0.3}^{+0.4}$ & $3.98_{-0.11}^{+0.32}$ & $4.95_{-0.17}^{+0.78}$ \\
$q_{\rm out}$ & $2.94^{+0.14}_{-0.09}$ & $2.88^{+0.14}_{-0.09}$ & $2.88^{+0.11}_{-0.15}$ & $2.84^{+0.14}_{-0.04}$
& $2.83_{-0.09}^{+0.13}$ & $2.56_{-0.11}^{+0.23}$ & $1.8_{-0.6}^{+0.3}$ & $2.23_{-0.59}^{+0.19}$ \\
$R_{\rm br}$ & $3.0_{-0.9}^{+0.4}$ & $3.17_{-0.24}^{+0.85}$ & $3.43_{-0.07}^{+0.55}$ & $3.50_{-0.05}^{+0.61}$ 
& $3.0_{-0.6}^{+0.8}$ & $6.4_{-1.1}^{+3.0}$ & $16.6_{-1.7}^{+13.7}$ & $7.1_{-0.5}^{+3.0}$ \\
$i$ [deg] & \multicolumn{4}{c}{$30.7^{+1.3}_{-1.6}$} & \multicolumn{4}{c}{$30.9^{+1.8}_{-2.3}$} \\
$a_*$ & \multicolumn{4}{c}{$0.958^{+0.009}_{-0.011}$} & \multicolumn{4}{c}{$0.0148^{+0.0014}_{-0.0048}$} \\
$\gamma$ & \multicolumn{4}{c}{--} & \multicolumn{4}{c}{$0.36_{-0.10}$} \\
$z$ & \multicolumn{4}{c}{$0.007749^\star$} & \multicolumn{4}{c}{$0.007749^\star$} \\
$\log\xi$ & $2.87^{+0.04}_{-0.06}$ & $2.988^{+0.022}_{-0.067}$ & $3.059^{+0.026}_{-0.013}$ & $3.134^{+0.011}_{-0.022}$ 
& $2.88_{-0.09}^{+0.04}$ & $2.96_{-0.08}^{+0.04}$ & $3.040_{-0.016}^{+0.011}$ & $3.116_{-0.027}^{+0.022}$ \\
$A_{\rm Fe}$ & \multicolumn{4}{c}{$3.09^{+0.24}_{-0.32}$} & \multicolumn{4}{c}{$3.13^{+0.21}_{-0.17}$} \\
$N_\text{\sc relxill\_nk}$~$(10^{-5})$ & $4.95^{+0.22}_{-0.34}$ & $6.04^{+0.28}_{-0.11}$ & $10.30^{+0.23}_{-0.55}$ & $12.9^{+0.6}_{-0.8}$ 
& $5.0_{-0.5}^{+0.3}$ & $5.9_{-0.5}^{+0.8}$ & $9.1_{-0.5}^{+0.8}$ & $11.8_{-1.4}^{+0.6}$ \\ 
\hline
{\sc xillver} &&&& \\
$\log\xi'$ & \multicolumn{4}{c}{$0^\star$} & \multicolumn{4}{c}{$0^\star$} \\
$N_\text{\sc xillver}$~$(10^{-5})$ & \multicolumn{4}{c}{$5.5^{+0.3}_{-0.5}$} & \multicolumn{4}{c}{$5.2^{+0.6}_{-0.6}$} \\
\hline
{\sc zgauss} &&&& &&&&\\
$E_{\rm line}$ [keV] & \multicolumn{4}{c}{$0.8143^{+0.0006}_{-0.0036}$} & \multicolumn{4}{c}{$0.8143^{+0.0033}_{-0.0006}$} \\
\hline
{\sc zgauss} &&&& &&&&\\
$E_{\rm line}$ [keV] & \multicolumn{4}{c}{$1.226^{+0.011}_{-0.008}$} & \multicolumn{4}{c}{$1.226^{+0.013}_{-0.008}$} \\
\hline
$\chi^2$/dof & \multicolumn{4}{c}{$3024.94/2683 = 1.12745$} & \multicolumn{4}{c}{$3022.84/2682 = 1.12708$} \\
\hline\hline
\end{tabular}
}
\vspace{0.2cm}
\caption{Summary of the best-fit values for the black hole model (left) and the wormhole model (right). For every model, we have four flux states: low flux state (1), medium flux state (2), high flux state (3), and very-high flux state (4). Note that some model parameters are supposed to keep the same value over different flux states, and are thus fitted as a single free parameter. The parameters that are thought to be able to vary over short timescales are fitted for every flux state. The ionization parameters ($\xi$, $\xi'$, $\xi_1$, and $\xi_2$) are in units erg~cm~s$^{-1}$. The reported uncertainties correspond to the 90\% confidence level for one relevant parameter. $^\star$ indicates that the parameter is frozen. See the text for more details. \label{t-fit}}
\end{table*}

\section{Analysis of MCG--6--30--15 \label{s-3}}

MCG--6--30--15 is a very bright Seyfert~1 galaxy at redshift $z = 0.007749$. Its spectrum is characterized by a broad and prominent iron K$\alpha$ line, and it is indeed the source in which a relativistically blurred iron line was clearly detected for the first time~\cite{tanaka95}. Among the many X-ray observations of MCG--6--30--15, we analyze those of \textsl{XMM-Newton} and \textsl{NuSTAR} in 2013~\cite{andrea}. In this way, we have both a high energy resolution around the iron line (thanks to the \textsl{XMM-Newton} data) and a broad energy band (thanks to the \textsl{NuSTAR} data up to 80~keV). The technical details concerning observations, data reduction, and data analysis are reported in Appendix~\ref{s-aaa}, while in this section we focus on the results.

In order to test the black hole and wormhole scenarios, we fit the data with a relativistic reflection model assuming the Kerr spacetime and our relativistic reflection model for traversable wormholes described in the previous section. For the Kerr model, we employ {\sc relxill}~\cite{jt1,jt2}. The reflection model for the wormhole spacetime is constructed within the {\sc relxill\_nk} package, so the two models only differ by the spacetime metric, while they share the same astrophysical model (morphology of accretion disk and corona) and non-relativistic reflection model (i.e. the reflection spectrum at the emission point in the rest-frame of the gas). In this way, we can compare the fits obtained from the two models and figure out if our observations can select one of the models and rule out the other one or, otherwise, if the quality of the fits is similar and our data cannot distinguish the two scenarios.

\begin{figure}[t]
\begin{center}
\includegraphics[type=pdf,ext=.pdf,read=.pdf,width=8.5cm,trim={1.0cm 2.5cm 0.5cm 0.5cm},clip]{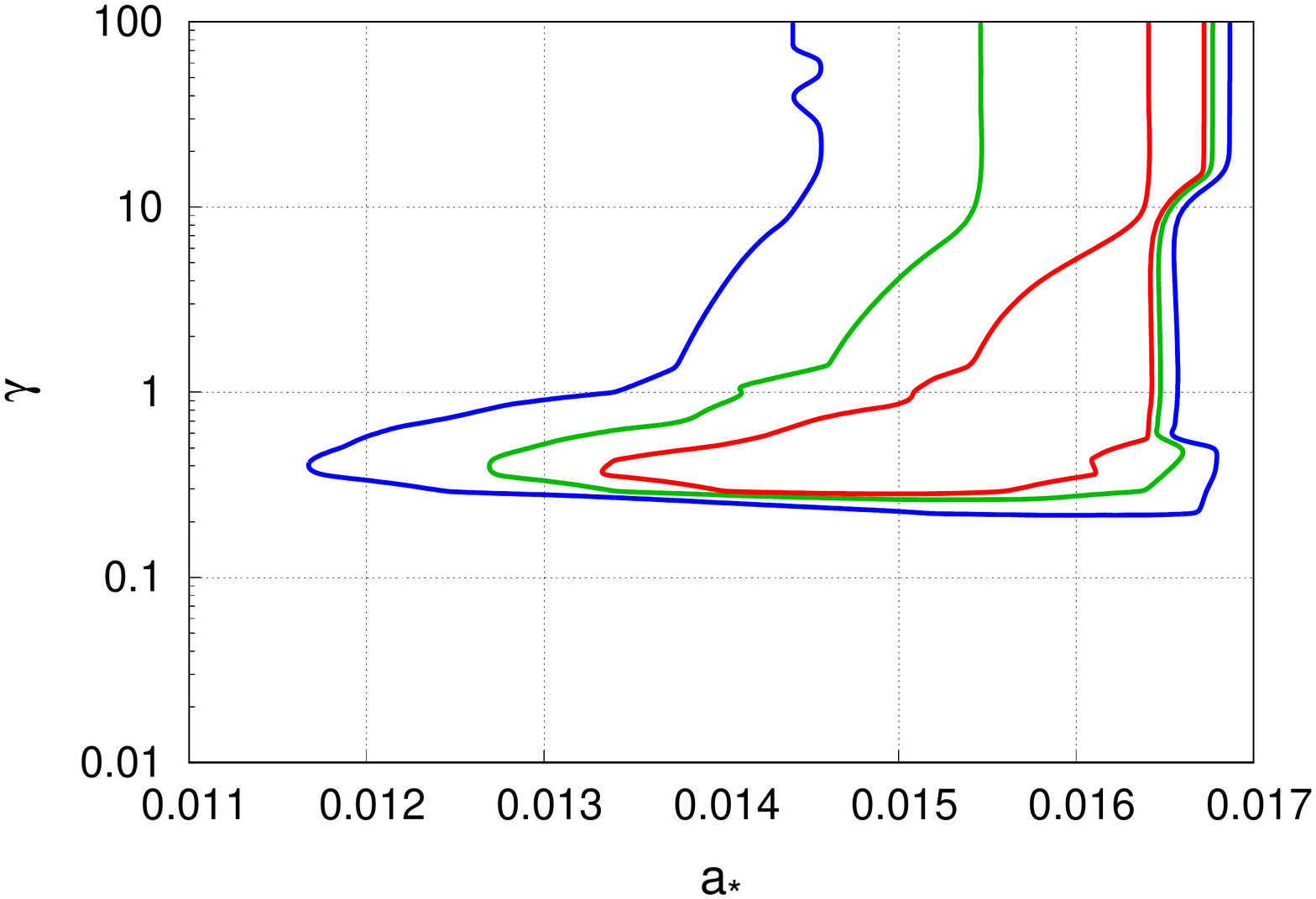}
\end{center}
\vspace{-0.4cm}
\caption{Constraints on the spin parameter $a_*$ and the parameter $\gamma$ from the fit of the wormhole model. The red, green, and blue curves correspond, respectively, to the 68\%, 90\%, and 99\% confidence level limits for two relevant parameters ($\Delta\chi^2 = 2.30$, 4.61, and 9.21, respectively). \label{f-res}}
\end{figure}

For the data analysis, we follow the study presented in Ref.~\cite{n3} and the technical details are reported in Appendix~\ref{s-aaa}. The source flux is very variable, so we group the data into four different flux states: low flux state, medium flux state, high flux state, and very-high flux state. We then fit all of the data sets together, bearing in mind that some model parameters must be the same over different flux states (e.g. spin of the compact object, inclination angle of the disk, iron abundance of the disk, etc.) because they are not supposed to vary over the timescales of the observations. Other model parameters are instead allowed to vary from one state to another (e.g. column densities of the warm absorbers, photon index of the coronal spectrum, ionization parameter of the disk, etc.) because they are expected to be able to vary over shorter time scales. The spectrum is characterized by a few components as already found in previous analyses of these observations~\cite{andrea,n3}, but the relativistic reflection component from the accretion disk is quite strong and can thus permit interesting constraints on the spacetime metric of the strong gravity region of the supermassive body at the center of MCG--6--30--15. The results of our fits for the black hole and the wormhole models are summarized in Tab.~\ref{t-fit}. The constraints on the spin parameter $a_*$ and the parameter $\gamma$ in the wormhole model are shown in Fig.~\ref{f-res}.

\section{Discussion and conclusions \label{s-4}}

The main result of our study to figure out whether current observations can distinguish the cases in which the supermassive objects in galactic nuclei, or at least the one at the center of MCG--6--30--15, is a black hole or a traversable wormhole is Tab.~\ref{t-fit}. This source is particularly suitable for this kind of test because it is very bright and its spectrum shows a prominent and broad iron line. We have fitted the simultaneous observations of \textsl{XMM-Newton} and \textsl{NuSTAR} in 2013 with the most advanced relativistic reflection models available today. In the first model, we assume that the central supermassive object is a Kerr black hole of general relativity. In the second model, we assume that the central object is a traversable wormhole whose metric is described by the line element in Eq.~(\ref{eq-metric}). The source flux is highly variable and there is some absorbing material along the line of sight, two ingredients that make the data analysis more challenging. On the other hand, the mass accretion rate should be around 30\% of the Eddington limit of the source~\cite{ref2}, so the deviations from the ideal thin accretion disk may be marginal~\cite{cfm,riaz}.

The minima of the $\chi^2$ in the black hole and wormhole models are comparable. The ratio plots between the data and the best-fit models of the two models are also similar and do not show clear unresolved features (see Figs.~\ref{f-rk} and \ref{f-rw} in Appendix~\ref{s-aaa}). The quality of the two fits is thus good and observations do not prefer one of the models over the other. While traversable wormholes would be definitively different objects with respect to black holes, our analysis cannot rule out the possibility that the supermassive objects at the center of every normal galaxy are wormholes instead of black holes. Considering the quality of the data of MCG--6--30--15 of 2013 from \textsl{XMM-Newton} and \textsl{NuSTAR}, we do not think that other sources or other observations with the available X-ray missions can say more on this issue. Much better data will only be possible with \textsl{Athena}, which is expected to be launched after 2030 and have an energy resolution near the iron line of 2.5~eV (compared to 150~eV of \textsl{XMM-Newton}).

Concerning tests with other techniques, interesting constraints may be obtained from imaging of supermassive black holes. The Event Horizon Telescope collaboration has recently released the image of the supermassive compact object at the center of the galaxy M87~\cite{Akiyama:2019cqa} and is expected to soon release a similar study for that at the center of the Galaxy. Observational data are consistent with their Kerr model; it would be interesting to check whether a model implementing our wormhole metric can provide a better or worse fit and, in the second case, whether the available data can already rule out the wormhole scenario. For examples, the wormholes in Ref.~\cite{p7} seem to be ruled out because the size of their shadow would be smaller than that observed for M87*~\cite{Akiyama:2019cqa,Akiyama:2019bqs}. A similar study could be done in the wormhole metric in Eq.~(\ref{eq-metric}) and considering the uncertainty on the mass of the black hole in M87. The detection of the shadow of SgrA*, the supermassive black hole in our Galaxy, will likely be a more stringent test, as its mass and distance are known with better precision. For the metric in Eq.~(\ref{eq-metric}) with $a = 0$, the radius of the shadow can be calculated to be $2.718$~$M$ (independent of $\gamma$), which is much smaller than $5.196$~$M$ expected for a Schwarzschild black hole, where $M$ is the mass of SgrA*~\cite{p7}.

Last, gravitational wave space antennas like \textsl{LISA} also promise to be able to test the nature of the supermassive objects in galactic nuclei with exquisite precision. They are expected to be launched after 2030.


\vspace{0.5cm}

{\bf Acknowledgments --}
This work was supported by the Innovation Program of the Shanghai Municipal Education Commission, Grant No.~2019-01-07-00-07-E00035, the National Natural Science Foundation of China (NSFC), Grant No.~11973019, and Fudan University, Grant No.~IDH1512060.


\appendix

\section{Data reduction and analysis \label{s-aaa}}

MCG--6--30--15 is a very bright Seyfert~1 galaxy. There have been many observations of this source over the past 25~years and with different X-ray missions. The source is characterized by a broad and prominent iron K$\alpha$ line, which makes MCG--6--30--15 a good candidate for testing the nature of the central supermassive object using X-ray reflection spectroscopy.

\begin{table}
\centering
\scalebox{1.0}{
\begin{tabular}{lcc}
\hline\hline
Mission & \hspace{0.2cm} Observation ID \hspace{0.2cm} & Exposure time (ks) \\
\hline
\textsl{NuSTAR} & 60001047002 & 23 \\
& 60001047003 & 127 \\
& 60001047005 & 30 \\
\hline
\textsl{XMM-Newton} & 0693781201 & 134 \\
& 0693781301 & 134 \\
& 0693781401 & 49 \\
\hline\hline
\end{tabular}
}
\vspace{0.2cm}
\caption{Basic details of the observations analyzed in the present work. \label{t-obs}}
\end{table}

\subsection{Observations}

Among the many X-ray observations of MCG--6--30--15, we chose those of \textsl{NuSTAR} and \textsl{XMM-Newton} in 2013. The observations with the two missions were simultaneous, started on 29~January~2013 and their total observational time is about 360~ks (\textsl{NuSTAR}) and 315~ks (\textsl{XMM-Newton}). The basic details of these observations are summarized in Tab.~\ref{t-obs}

\subsection{Data reduction}

\textsl{NuSTAR} is equipped with two coaligned telescopes with focal plane modules, called FPMA and FPMB~\cite{nustar}. For the level~1 data products, we used the NuSTAR Data Analysis Software (NUSTARDAS). The downloaded raw data were then converted to event files (level~2 products) by using the HEASOFT task NUPIPELINE and the latest calibration data file from the NuSTAR calibration database (CALDB), version 20180312. For the source events, we took a circular region centered at the source of radius 70~arcsec. For the background events, we took a circular region on the same CCD of radius 100~arcsec. Light curves and spectra were extracted using the event and region files by running the NUPRODUCTS task. Spectra were rebinned to a minimum of 70~counts per bin in order to apply the $\chi^2$ statistics.

\textsl{XMM-Newton} is equipped with three X-ray CCD cameras, two EPIC-MOS and one EPIC-Pn camera~\cite{xmm}. In the observations of MCG--6--30--15, they operated in medium filter and small window modes. In our analysis, we only used the EPIC-Pn data, while we ignored the EPIC-MOS data because they are strongly affected by pile-up. We used SAS version~16.0.0 to convert the raw data into event files, which were then combined into a single FITS file using the ftool FMERGE. We used TABTIGEN to generate good time intervals (GTIs), which were then employed to filter the event files. For the source events, we took a circular region centered at the source of radius 40~arcsec. For the background events, we took a circular region of radius 50~arcsec. Response files were produced after backscaling. Spectra were rebinned to oversample the instrumental resolution by at least a factor of 3 and to have at least 50~counts in each background-subtracted bin.

\begin{figure}[t]
\vspace{-1.0cm}
\begin{center}
\includegraphics[width=9cm,trim={0.5cm 0 3cm 7cm},clip]{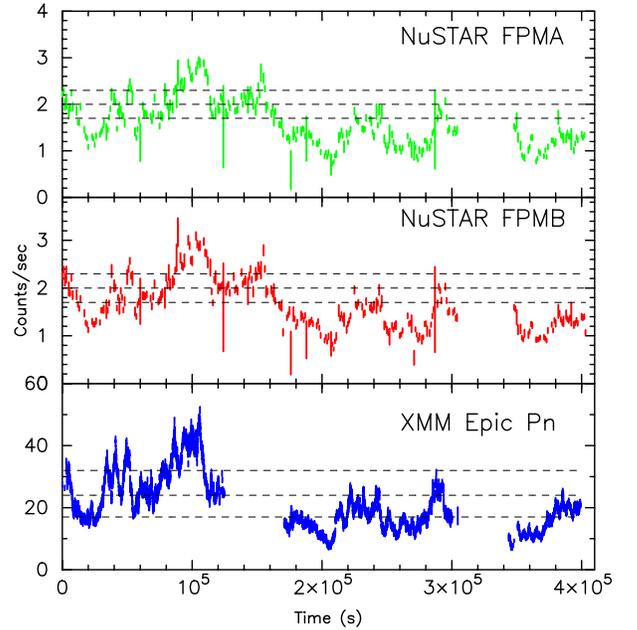}
\end{center}
\vspace{-0.7cm}
\caption{Light curves and flux states for \textsl{XMM-Newton}/EPIC-Pn (0.5-10~keV) and \textsl{NuSTAR}/FPMA and \textsl{NuSTAR}/FPMB (3-80~keV). The three dashed horizontal lines separate the four different flux states. \label{f-lc}}
\end{figure}

\subsection{Data analysis}

MCG--6--30--15 is a very variable source in the X-ray band. We used the ftool mgtime to find the common GTIs of the two telescopes and ensure to work with simultaneous data. We grouped the data into four data sets according to the flux state of the source: low flux state, medium flux state, high flux state, and very-high flux state. Eventually we got 12~data sets, namely 4~flux states for 3~instruments (\textsl{NuSTAR}/FPMA, \textsl{NuSTAR}/FPMB, \textsl{XMM-Newton}/EPIC-Pn). The four flux state data sets were created in such a way that the spectral data counts are similar in every flux state. Fig.~\ref{f-lc} shows the light curves and the flux state separation. The same grouping scheme was used in Ref.~\cite{n3}, while a different grouping scheme based on the hardness of the source was used in Ref.~\cite{andrea}.

We combined the \textsl{XMM-Newton} and \textsl{NuSTAR} data by freezing the normalization constant of \textsl{XMM-Newton} to 1 and leaving free those of \textsl{NuSTAR}/FPMA and \textsl{NuSTAR}/FPMB. After the fits, we checked that the ratios between the constants of \textsl{NuSTAR}/FPMA and \textsl{NuSTAR}/FPMB were between 0.95 and 1.05 for every flux state.

In the \textsl{XMM-Newton}/EPIC-Pn data, we ignored the energy range 1.5-2.5~keV because of the presence of a Gaussian around 2~keV, which is interpreted as an effect of the golden edge in the response file due to miscalibration in the long-term charge transfer inefficiency (CTI); that is, how photon energies are reconstructed after detection. More details can be found in Ref.~\cite{andrea}.

\begin{figure*}[t]
\begin{center}
\vspace{-0.6cm}
\includegraphics[type=pdf,ext=.pdf,read=.pdf,width=8.5cm,trim={1.5cm 0.5cm 2.5cm 17cm},clip]{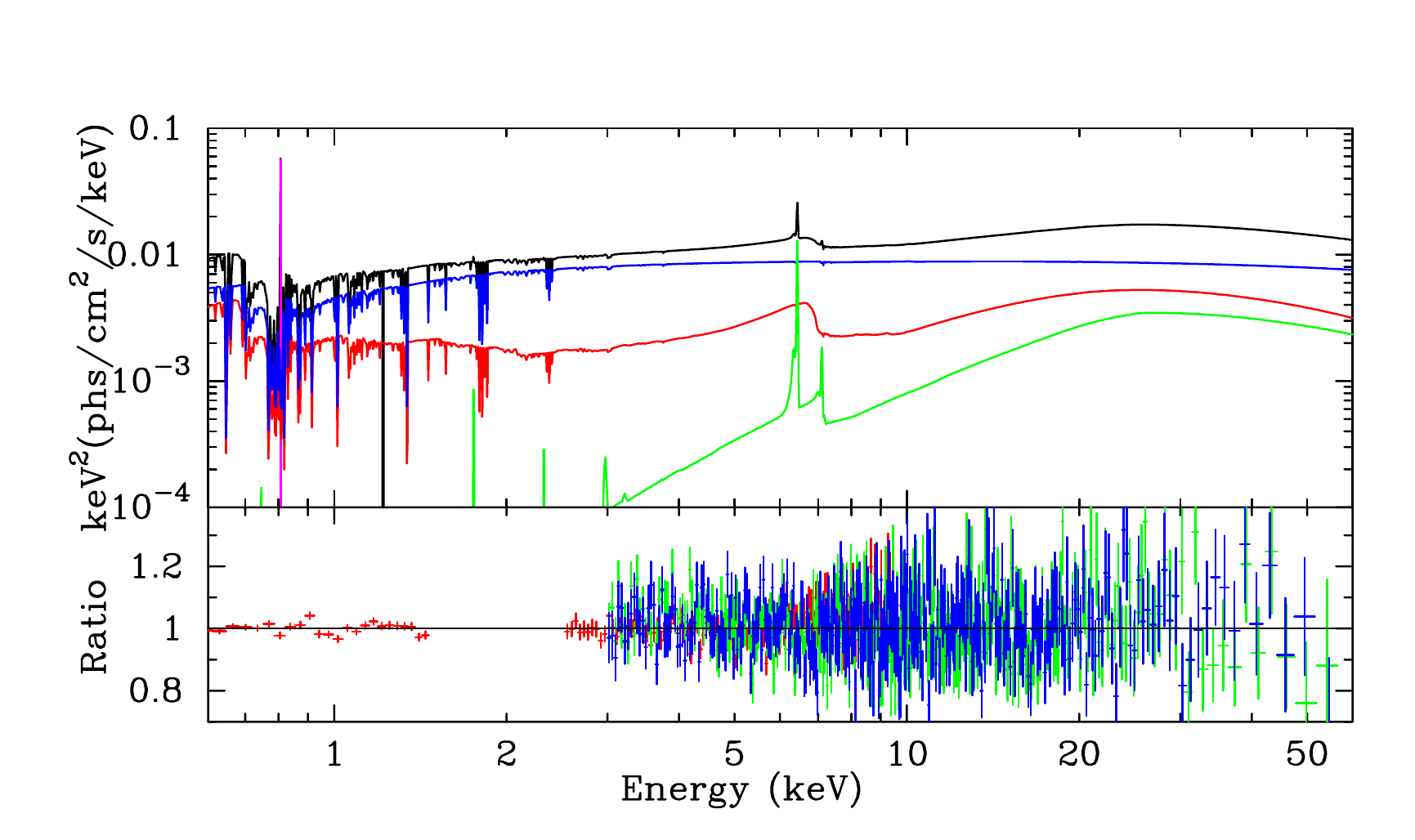}
\includegraphics[type=pdf,ext=.pdf,read=.pdf,width=8.5cm,trim={1.5cm 0.5cm 2.5cm 17cm},clip]{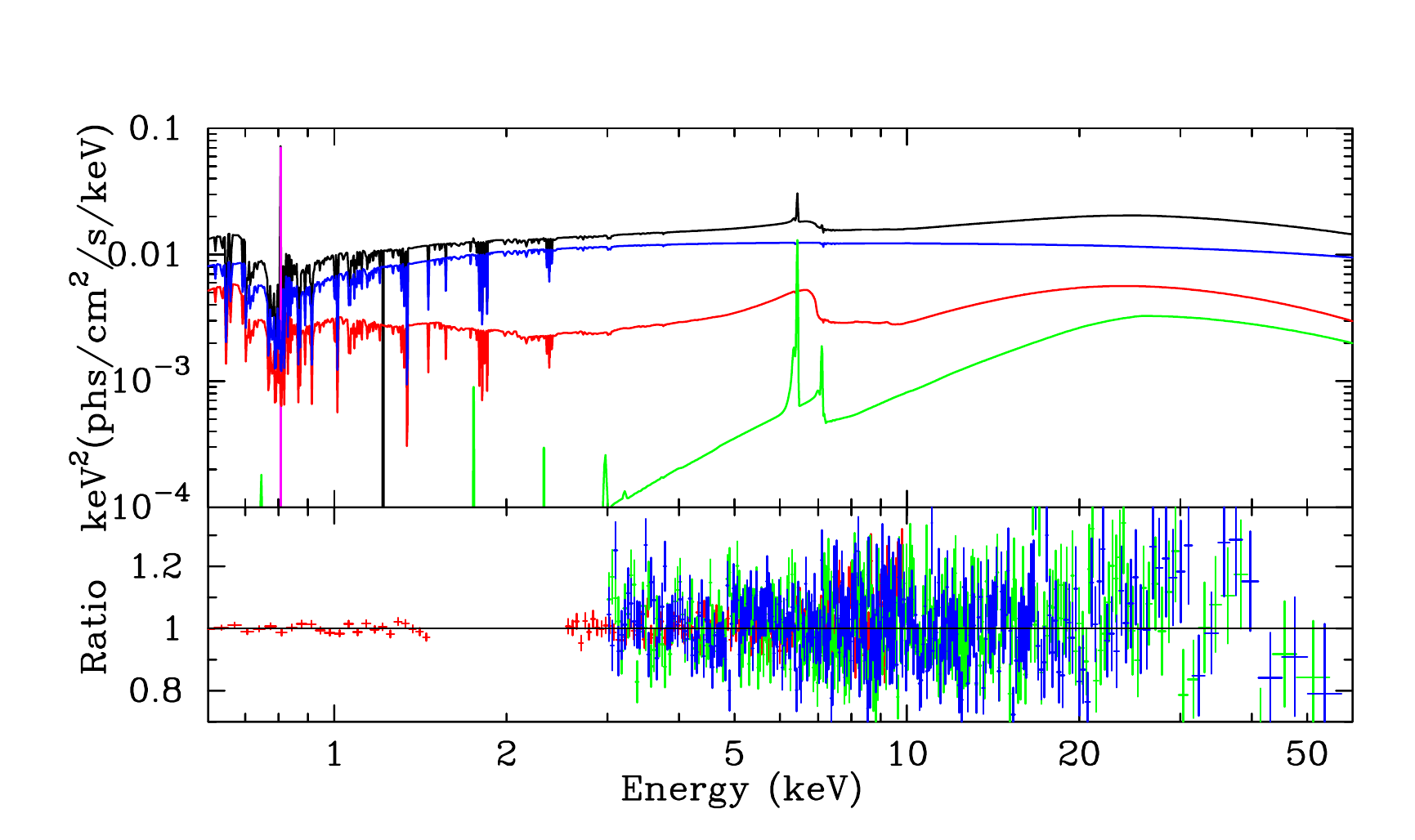}\\
\vspace{-0.4cm}
\includegraphics[type=pdf,ext=.pdf,read=.pdf,width=8.5cm,trim={1.5cm 0.5cm 2.5cm 17cm},clip]{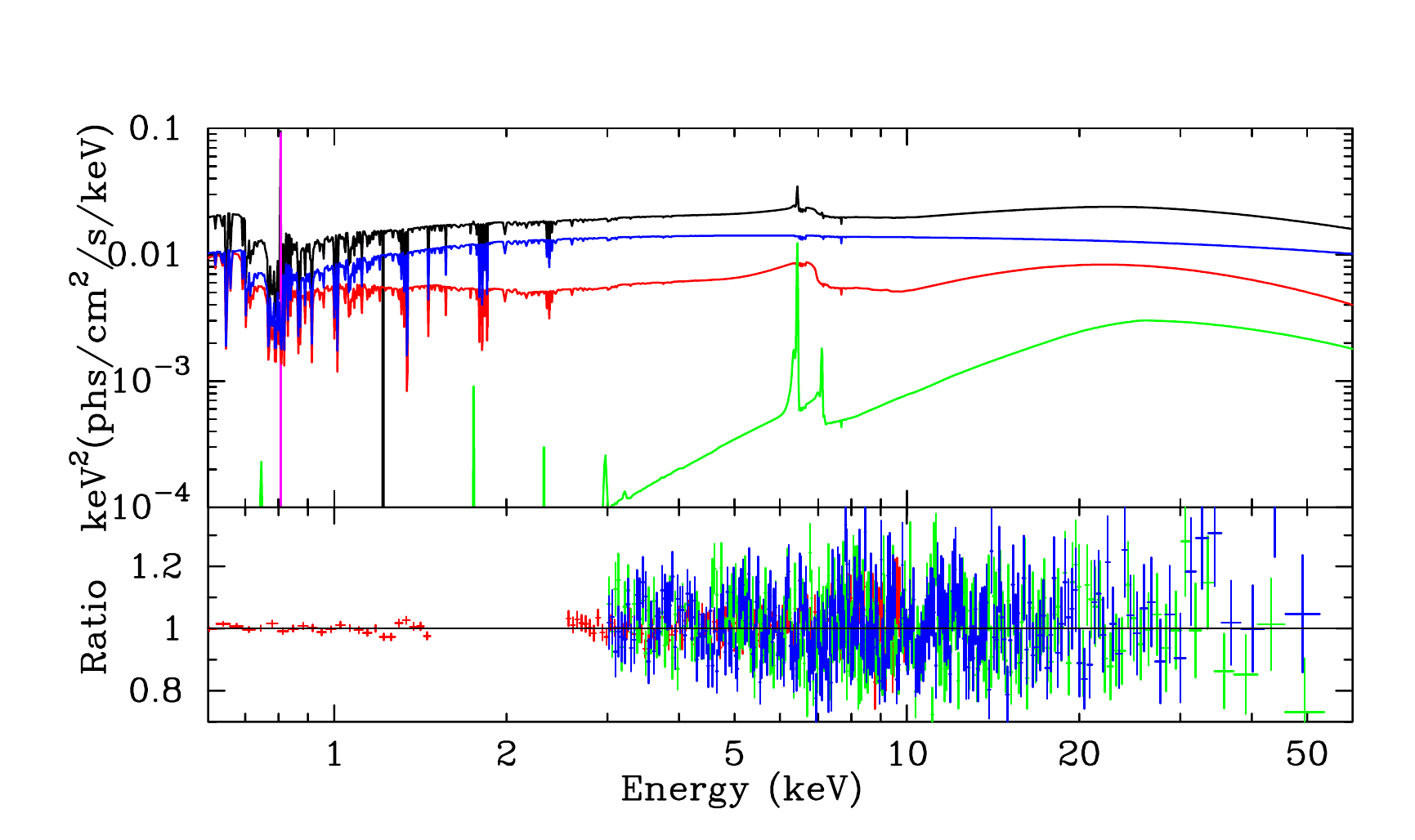}
\includegraphics[type=pdf,ext=.pdf,read=.pdf,width=8.5cm,trim={1.5cm 0.5cm 2.5cm 17cm},clip]{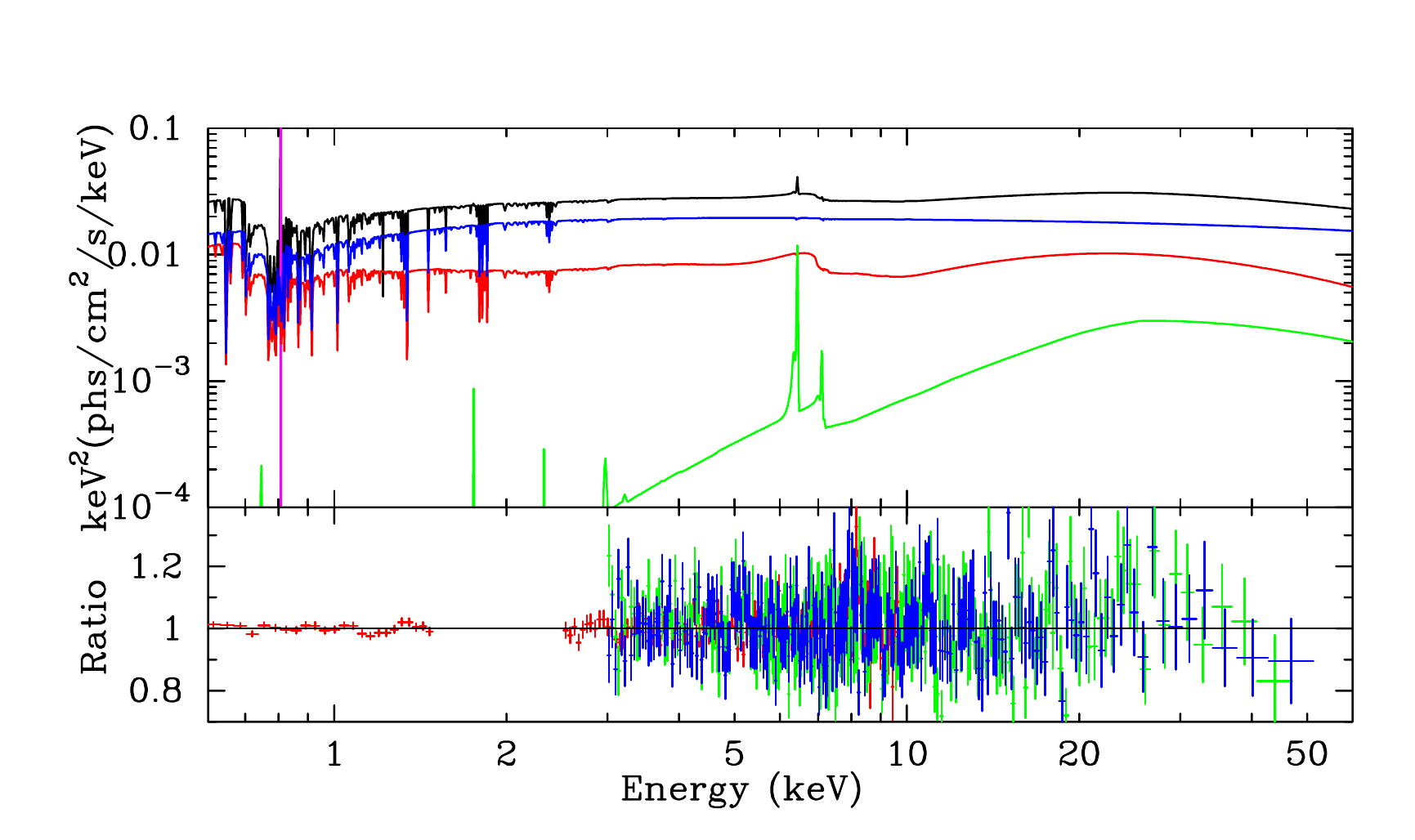}
\end{center}
\vspace{-0.5cm}
\caption{Black hole model -- Spectra of the best-fit models and ratios between the data and the best-fit model for the low flux state (top left panel), the medium flux state (top right panel), the high flux state (bottom left panel), and the very-high flux state (bottom right panel). In the spectra, the black curve is for the total spectrum, the blue curve is for the coronal component ({\sc cutoffpl}), the red curve is for the relativistic reflection component ({\sc relxill}), the green curve is for the non-relativistic reflection component ({\sc xillver}), and the magenta curve is for narrow emission lines ({\sc zgauss}). In the ratio plots, the red crosses are used for \textsl{XMM-Newton}, the green crosses for \textsl{NuSTAR}/FPMA, and the blue crosses for \textsl{NuSTAR}/FPMB. \label{f-rk}}
\vspace{-0.3cm}
\begin{center}
\includegraphics[type=pdf,ext=.pdf,read=.pdf,width=8.5cm,trim={1.5cm 0.5cm 2.5cm 17cm},clip]{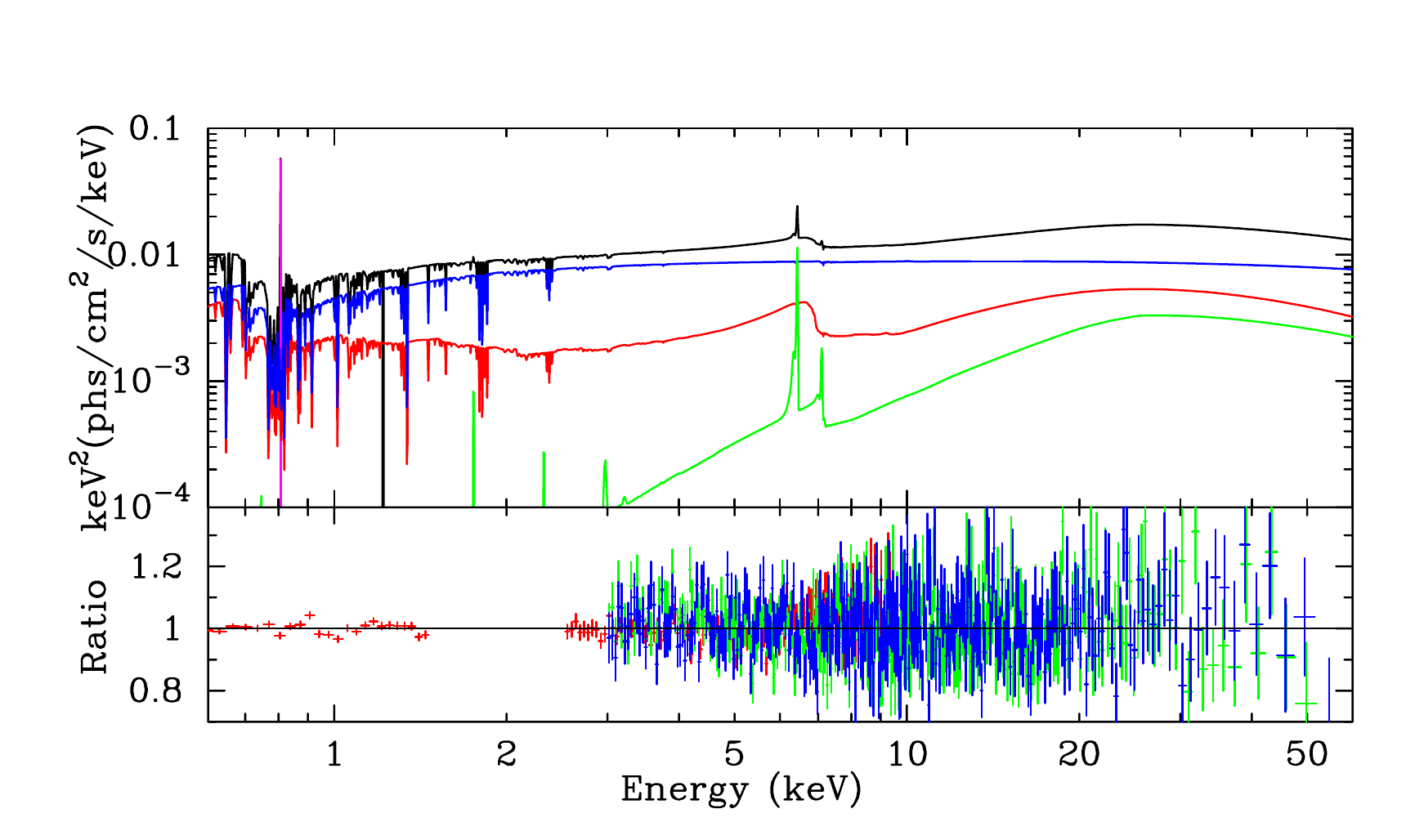}
\includegraphics[type=pdf,ext=.pdf,read=.pdf,width=8.5cm,trim={1.5cm 0.5cm 2.5cm 17cm},clip]{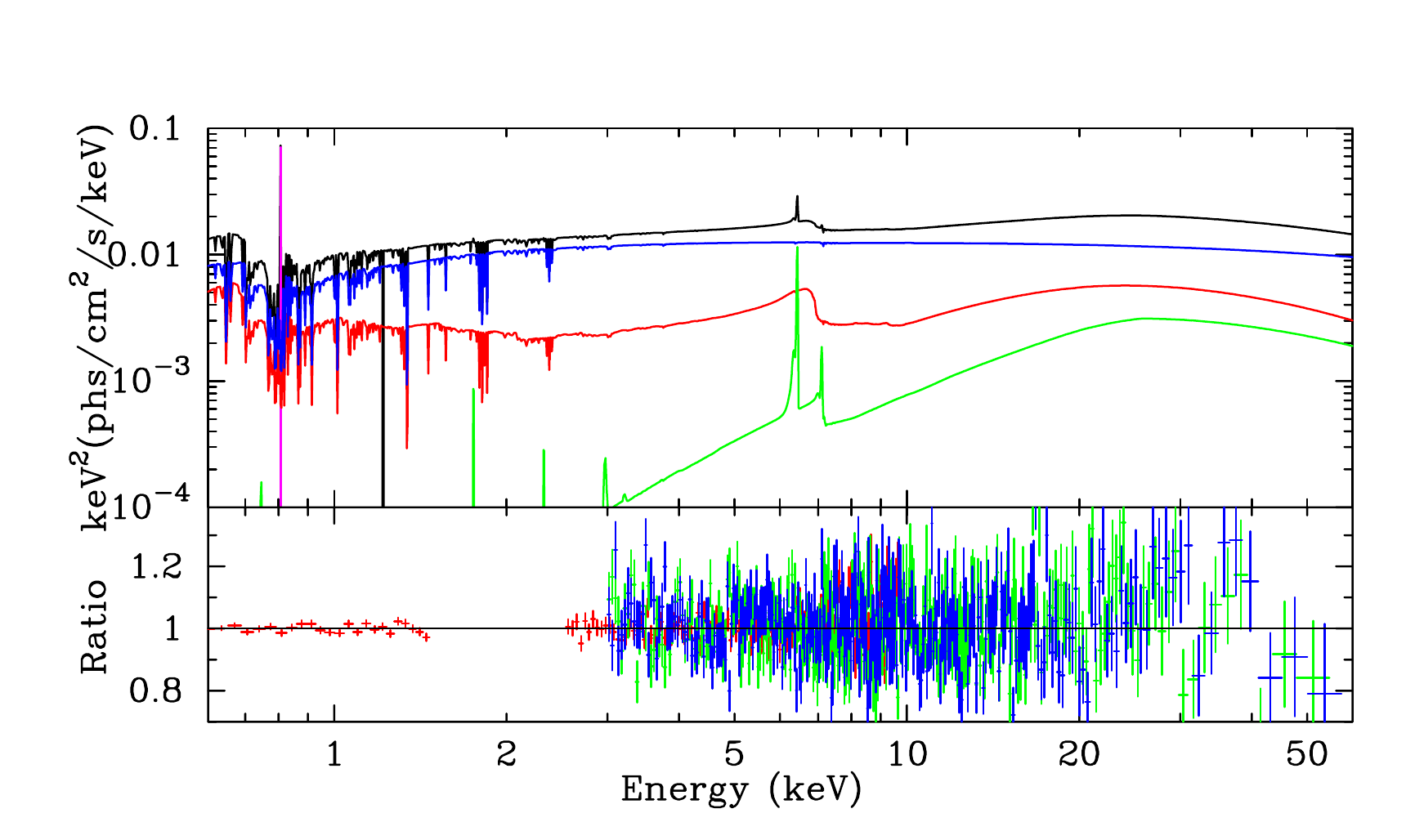}\\
\vspace{-0.4cm}
\includegraphics[type=pdf,ext=.pdf,read=.pdf,width=8.5cm,trim={1.5cm 0.5cm 2.5cm 17cm},clip]{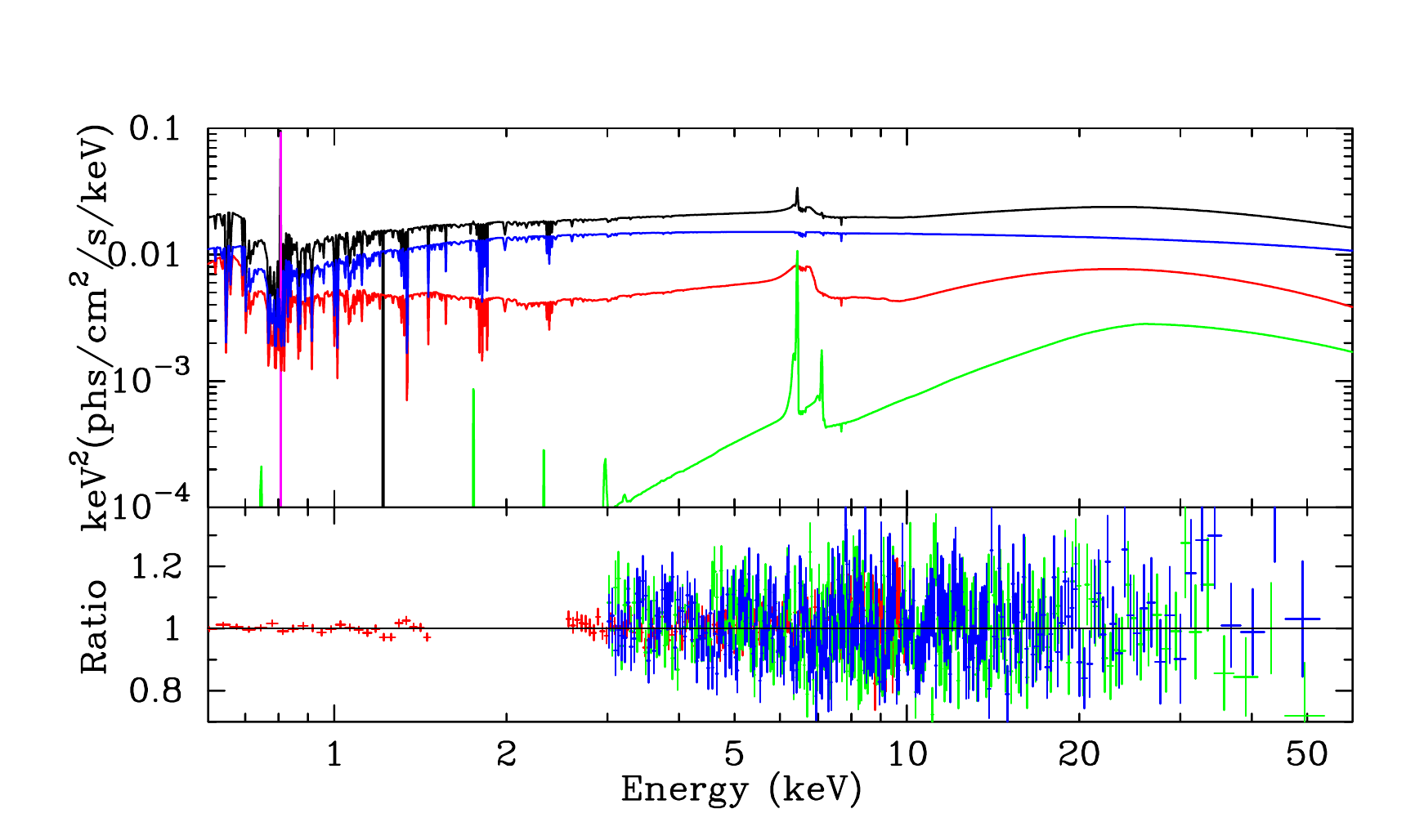}
\includegraphics[type=pdf,ext=.pdf,read=.pdf,width=8.5cm,trim={1.5cm 0.5cm 2.5cm 17cm},clip]{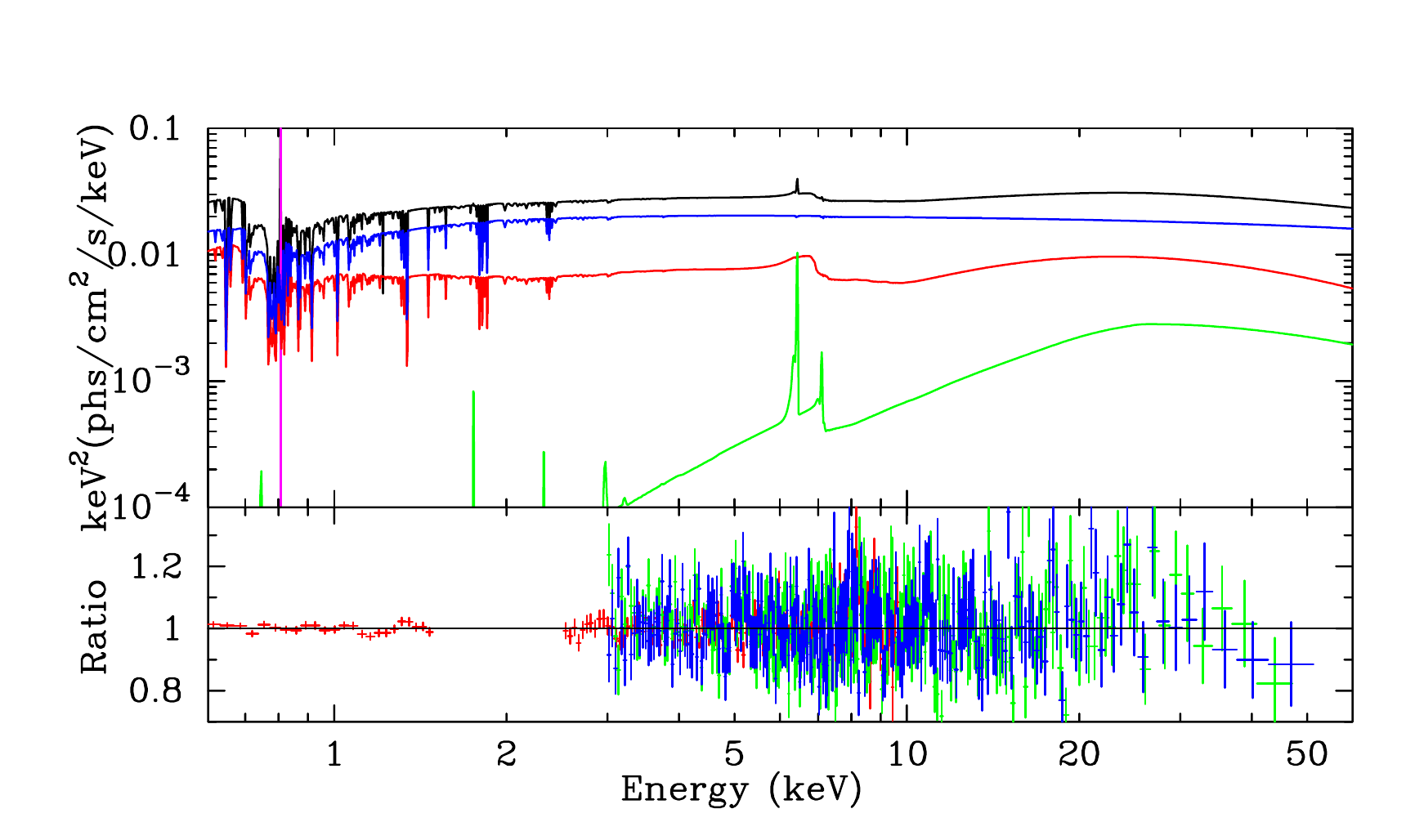}
\end{center}
\vspace{-0.5cm}
\caption{Wormhole model -- Spectra of the best-fit models and ratios between the data and the best-fit model for the low flux state (top left panel), the medium flux state (top right panel), the high flux state (bottom left panel), and the very-high flux state (bottom right panel). In the spectra, the black curve is for the total spectrum, the blue curve is for the coronal component ({\sc cutoffpl}), the red curve is for the relativistic reflection component (see Section~\ref{s-2}), the green curve is for the non-relativistic reflection component ({\sc xillver}), and the magenta curve is for narrow emission lines ({\sc zgauss}). In the ratio plots, the red crosses are used for \textsl{XMM-Newton}, the green crosses for \textsl{NuSTAR}/FPMA, and the blue crosses for \textsl{NuSTAR}/FPMB. \label{f-rw}}
\end{figure*}

We fitted the four flux state data sets together. The model parameters that are supposed to be constant over different flux states (spin parameter, viewing angle, iron abundance, etc.) were fitted assuming that the value of the parameter was the same for all flux states. The model parameters that can vary over short timescales were left free for every flux state. The model to fit these observations was already discussed in Refs.~\cite{andrea,n3}. In XSPEC language~\cite{xspec}, the final model is {\sc tbabs$\times$warmabs$_1$$\times$warmabs$_2$$\times$dustyabs$\times$(cutoffpl + refl + xillver + zgauss + zgauss)}. {\sc tbabs} describes the Galactic absorption and we froze the column density to $N_{\rm H} = 3.9 \cdot 10^{22}$~cm$^{-2}$~\cite{dl90}. {\sc warmabs$_1$} and {\sc warmabs$_1$} describe two ionized absorbers; their tables were generated with {\sc xstar} version~2.41. {\sc dustyabs} describes a neutral absorber due to the presence of dust around the source and only modifies the soft X-ray band~\cite{l00}. {\sc cutoffpl} describes the spectrum of the corona, modeled by a power-law with an exponential cutoff energy. {\sc refl} describes the relativistic reflection spectrum from the accretion disk: for the Kerr scenario we used {\sc relxill}~\cite{jt1,jt2}, while for the wormhole scenario we used the model described in Section~\ref{s-2}. There are 8 free parameters in the black hole model and 9 free parameters in the wormhole one. The emissivity profile is modeled with a broken power law, so we have the inner emissivity index ($q_{\rm in}$), the outer emissivity index ($q_{\rm out}$), and the breaking radius ($R_{\rm br}$). The other free parameters are: the inclination angle of the disk with respect to the line of sight of the observer ($i$), the spin parameter ($a_*$), the ionization parameter ($\xi$), the iron abundance ($A_{\rm Fe}$), the normalization of the component ($N_\text{\sc relxill\_nk}$), and, in the case of the wormhole, the parameter $\gamma$. {\sc xillver} describes a non-relativistic reflection component due to some cold material at larger distance~\cite{xill1,xill2}. {\sc zgauss} is used to describe a narrow oxygen line around 0.81~keV and a narrow absorption feature around 1.22~keV. The latter may be interpreted as a blueshifted oxygen absorption from some relativistic outflow. Fig.~\ref{f-rk} and Fig.~\ref{f-rw} show the spectra of the best-fit models and the ratios between the data and the best-fit models for the four flux states and, respectively, the black hole and wormhole scenarios. Tab.~\ref{t-fit} shows the best-fit values of the two scenarios.

\end{document}